\documentclass[11pt,oneside,english]{amsart}
\usepackage{geometry}
\usepackage{amsmath,amsfonts, dsfont}
\usepackage{amssymb}
\usepackage{tikz}
\usepackage[T1]{fontenc}
\usepackage[latin9]{inputenc}
\usepackage{color}
\usepackage{latexsym}
\usepackage{amsbsy}
\usepackage{amstext}
\usepackage{graphicx}
\usepackage{amsmath}
\usepackage{amssymb}
\usepackage{tikz}

\usepackage{amsthm}
\usepackage{amstext}
\usepackage{amssymb}
\usepackage{graphicx}
\usepackage{esint}
\usepackage{babel}

\usepackage{times}
\usepackage{color}
\usepackage{fancyhdr}
\AtBeginDocument{

}
\begin{document}
\global\long\def\bK{\mathbb{K}}
 \global\long\def\bC{\mathbb{C}}
 \global\long\def\bR{\mathbb{R}}
 \global\long\def\bZ{\mathbb{Z}}
 \global\long\def\bN{\mathbb{N}}
 \global\long\def\bQ{\mathbb{Q}}
 \global\long\def\bH{\mathbb{H}}

 \global\long\def\half{\frac{1}{2}}
 \global\long\def\ii{\mathrm{i}}

 \global\long\def\bdry{\partial}
 \global\long\def\cl#1{\overline{#1}}

\global\long\def\PR{\mathsf{P}}
 \global\long\def\EX{\mathsf{E}}
 \global\long\def\sU{\mathcal{U}}

\global\long\def\GL{\mathrm{GL}}
 \global\long\def\SL{\mathrm{SL}}
 \global\long\def\gl{\mathfrak{gl}}
 \global\long\def\sl{\mathfrak{sl}}
 \global\long\def\TL{\mathrm{TL}}

\global\long\def\Hom{\mathrm{Hom}}
 \global\long\def\End{\mathrm{End}}
 \global\long\def\Aut{\mathrm{Aut}}
 \global\long\def\Rad{\mathrm{Rad}}
 \global\long\def\Ext{\mathrm{Ext}}

\global\long\def\Kern{\mathrm{Ker}}
 \global\long\def\Imag{\mathrm{Im}}

\global\long\def\Alt{\mathrm{Alt}}

\global\long\def\dmn{\mathrm{dim}}
 \global\long\def\spn{\mathrm{span}}
 \global\long\def\tens{\otimes}
 \global\long\def\Mat{\mathrm{Mat}}
 \global\long\def\unitmat{\mathbb{I}}
 \global\long\def\id{\mathrm{id}}
 \global\long\def\diag{\mathrm{diag}}
 \global\long\def\unit{\mathbf{1}}

\global\long\def\set#1{\left\{  #1\right\}  }
 \global\long\def\sgn{\mathrm{sgn} }

\global\long\def\re{\Re\mathfrak{e}}
 \global\long\def\im{\Im\mathfrak{m}}
 \global\long\def\arg{\mathrm{arg}}
 \global\long\def\isom{\cong}

\global\long\def\op{\mathrm{op}}
 \global\long\def\cop{\mathrm{cop}}

\global\long\def\eps{\varepsilon}
 \global\long\def\const{\mathrm{const.}}

\global\long\def\binomial#1#2{{#1  \choose #2}}

\global\long\def\Hspace{\mathcal{F}}
 \global\long\def\Hinner#1#2{\left\langle #1,\,#2\right\rangle _{\Hspace}}

\global\long\def\spin{\sigma}
 \global\long\def\spinconf{\boldsymbol{\sigma}}
 \global\long\def\energy{\varepsilon}
 \global\long\def\contourset{\mathcal{C}}

\global\long\def\rect{R}
 \global\long\def\rectTop{\mathrm{top}}
 \global\long\def\rectBot{\mathrm{bot}}

\global\long\def\NE{\mathrm{NE}}
 \global\long\def\NW{\mathrm{NW}}
 \global\long\def\SW{\mathrm{SW}}
 \global\long\def\SE{\mathrm{SE}}

\global\long\def\iinterval#1{\llbracket\,\!#1\,\!\rrbracket}
 \global\long\def\hiinterval#1{\llbracket\,\!#1\,\!\rrbracket^{*}}

\global\long\def\Srow{\mathrm{Srow}}
 \global\long\def\Cliff{\mathrm{Cliff}}
 \global\long\def\CliffGen{\mathrm{Cliff}^{(1)}}
 \global\long\def\RH{\mathrm{RH}}
 \global\long\def\sP{\mathcal{P}}

\global\long\def\SLE{\mathrm{SLE}}
 \global\long\def\SLEk{\mathrm{SLE}_{\kappa}}

\global\long\def\iinterval#1{\llbracket\,\!#1\,\!\rrbracket}
 \global\long\def\hiinterval#1{\llbracket\,\!#1\,\!\rrbracket^{*}}

 \global\long\def\Srow{\mathrm{Srow}}
 \global\long\def\Cliff{\mathrm{Cliff}}
 \global\long\def\CliffGen{\mathrm{Cliff}^{(1)}}
 \global\long\def\RH{\mathrm{RH}}
 \global\long\def\sP{\mathcal{P}}

\numberwithin{equation}{section}
\title{Conformal field theory, vertex operator algebra and stochastic Loewner evolution in Ising model}
\author{Ali Zahabi} 
\begin{abstract}
We review the algebraic and analytic aspects of the conformal field theory (CFT) and its relation to the stochastic Loewner evolution (SLE) in an example of the Ising model. We obtain the scaling limit of the correlation functions of Ising free fermions on an arbitrary simply connected two-dimensional domain $D$. Then, we study the analytic and algebraic aspects of the fermionic CFT on $D$, using the Fock space formalism of fields, and the Clifford vertex operator algebra (VOA). These constructions lead to the conformal field theory of the Fock space fields and the fermionic Fock space of states and their relations in case of the Ising free fermions. Furthermore, we investigate the conformal structure of the fermionic Fock space fields and the Clifford VOA, namely the operator product expansions, correlation functions and differential equations. Finally, by using the Clifford VOA and the fermionic CFT, we investigate a rigorous realization of the $CFT/SLE$ correspondence in the Ising model. First, by studying the relation between the operator formalism in the Clifford VOA and the SLE martingale generators, we find an explicit Fock space for the SLE martingale generators. Second, we obtain a subset of the SLE martingale observables in terms of the correlation functions of fermionic Fock space fields which are constructed from the Clifford VOA.
\end{abstract}
\maketitle
\section{Introduction}
The Ising model was introduced and studied in 1925 by W. Lenz and E. Ising as a model describing ferromagnets on lattice. The Ising model consists of spins $\sigma_{\alpha}=\pm1$, on the vertices $\alpha$ of the lattice, interacting by short range neighborhood self-interactions as well as interactions with the external magnetic field $B$. In the case of $B=0$, in contrast to the one-dimensional model, two-dimensional Ising model in $\mathbb{Z}^2$ possesses a second order phase transition at critical inverse temperature ($\beta= \frac{1}{k_BT}$), $\beta_c=\frac{1}{2}\ln {(\sqrt{2}-1)}$, as shown in the pioneering works by Peierls, Kramers, Wannier and Onsager, \cite{Pe36}, \cite{KrWa41}, \cite{Ons44} and \cite{KaOn49}. At $\beta=\beta_c$ specific heat and magnetic susceptibility diverge to infinity and for $\beta > \beta_c$ spontaneous breaking of the symmetry leads to a nonzero magnetization. Moreover, due to infinite dimensional symmetry, many physical properties of the $2d$ Ising model such as free energy and spin correlations at $B=0$ can be computed exactly.

The transfer matrix formalism is one of the approaches toward the exact results in Ising model, \cite{Ba08}. Specially the free fermion operators play an essential role in this formalism, \cite{Kau49}. Moreover, the Fock space representations of the transfer matrix formalism and free fermions have been studied extensively, for a review see \cite{Pal07}.
In this paper, we study the connections between the discrete fermionic formalism of the Ising model and the rigorous aspects of a fermionic conformal field theory that describes the scaling limit of the model.

There is a common belief that the scaling limit of the lattice models such as Ising model at criticality are described by a field theory with the conformal symmetry. However, it is known that the spin operator is not enough to describe the continuum limit of the theory. In fact, it is believed that the critical Ising model in the continuum limit is described by a conformal field theory, namely the theory of free fermionic fields. Roughly speaking, the fermionic field is identified with the scaling limit of the fermionic operator on the lattice. However, there were no exact proofs about the scaling limit of the Ising model free fermions and their correlation functions.
Recently, the rigorous methods from discrete analysis and probability theory have provided exact proofs about the conformal invariance in the scaling limit of the Ising model at criticality, for a good review on general aspects of discrete holomorphicity see \cite{Car09} and \cite{DuSm11}. We have used these techniques to obtain the scaling limit of the correlation functions of Ising free fermions with specific boundary conditions. The continuum correlation functions are obtained from the scaling limit of the lattice correlation functions, \cite{HKZ12}. By means of these methods, we find a proof of the Pfaffian formula for the correlation functions of free fermionic fields, in the scaling limit.

Moreover, a new rigorous formulation of the continuum Fock space of fields and their properties; CFTs on domains with boundaries in the case of Gaussian free fields is proposed in \cite{KaMa11}. 
We have extended and adopted a similar formulation to obtain a conformal field theory on bounded domains in the case of free fermion fields of the Ising model. In this approach, we have obtained the characteristic features of fermionic conformal field theory on a bounded domain such as transformation rules for fields and their correlation functions, the operator product expansion of fields, Virasoro algebra representation and the Ward identity.

On the other hand, vertex operator algebra (VOA) provides a concrete mathematical language for CFT, \cite{Ka98}. The vertex operator algebra is an algebraic construction for conformal field theory in terms of formal power series.
The general VOA has been adopted in different cases for different purposes such as VOA for bosonic and fermionic fields. In this paper we have used the Clifford VOA for fermionic fields which has the Clifford algebra symmetry in addition to Virasoro algebra symmetry. The Clifford VOA, as an equivalent algebraic formalism to fermionic conformal field theory, turns out to be useful in study of scaling limit of the Ising model at criticality. Specially, we have obtained the Fock space of fermionic states in terms of VOA vector space.

From a different perspective, Stochastic Loewner evolution plays a crucial role in this picture. The SLE is a stochastic process that is defined by a stochastic differential equation, the Loewner equation with the Brownian motion as a driving force. In general, SLE curves explain the scaling limit of the interfaces of the statistical lattice models on domains with boundary, at critical temperature. Specially, it has been proved in \cite{CDHKS12}, that the scaling limit of the interfaces in $2d$ critical Ising model is described by a Schramm Loewner evolution, $SLE_3$. In SLE, the probability measures of the interface curves satisfy the conformal symmetry and the Markov property. These are physically expected conditions that the scaling limit of the interfaces should satisfy.

In this paper we combine the approach of Clifford VOA and the Fock space of conformal fermionic fields in order to obtain a unified picture of a conformal field theory describing the scaling limit of the critical Ising model. To have a unified picture, we need a mapping between the Clifford VOA and correlation functions of the Fock space fields which satisfies the axioms that are reflecting the analytic and algebraic aspects of the underlying conformal symmetry in the scaling limit of Ising model at criticality. This has been done through the main theorem of this paper.

Another aspect of this study refers to the well-known CFT/SLE correspondence, \cite{BaBe06}. We employ the aforementioned framework of the VOA and Fock space fields approach to CFT, and their interrelations in order to concretely investigate an example of the CFT/SLE correspondence in the case of Ising model. We have obtained the results indicating a rigorous realization of the fermionic CFT/$SLE_3$ in terms of Clifford VOA and Fock space of fermionic conformal fields. The Clifford VOA provides a fermionic Fock space for the $SLE_3$ martingale generators and furthermore, a large collection of $SLE_3$ martingale observables are explicitly written in terms of the correlation functions of the Fock space fields, i.e. corresponding fields to the states of the Clifford VOA.

This paper is organized as follows. In section two, we review the transfer matrix formalism for Ising free fermions and we obtain the scaling limit of the lattice correlation functions by using discrete holomorphicity techniques. In section three, we study the fermionic CFT on an arbitrary bounded domain and operator product expansions, correlation functions and differential equations. In section four, the Clifford VOA is reviewed and its relation to fermionic CFT is studied. In section five, by using the results from sections three and four, the CFT/SLE correspondence is explicitly reviewed in the Ising model example.
\section{Fermionic theory of Ising model}
In this section we first explain the transfer matrix formalism which is an approach towards an exact solution of two-dimensional Ising model on a rectangle with specific boundary conditions. Our aim in this section is to define and compute the correlation functions of any operator. Then, we introduce the notion of free fermions in the Ising model and compute their correlation functions on the lattice.
Eventually, by using methods from discrete complex analysis, the scaling limit of the fermionic correlation functions on the rectangular lattice are obtained. We will observe that these are CFT correlation functions on the half-plane, in the Fock space construction of the free fermionic fields.
\subsection{Transfer matrix formalism}
The Ising model on the domain $\Lambda= \{(j,i)\in \mathbb{Z}^2\mid |j|\leq M, |i|\leq N\}$, consists of spins $\sigma_{\alpha}=\pm1$, on the vertices $\alpha$ of the lattice in the domain $\Lambda$. The model is parameterized by the inverse temperature $\beta$ and nearest neighbor interaction coupling $J$ between $<\alpha,\alpha'>$, i.e. the pairs of sites that are nearest neighbors. The Ising model is defined by its partition function,
\begin{equation}\label{Ising partition function}
Z_\Lambda(\beta) = \sum_{\sigma\in \mathcal{C}_\Lambda}\exp{\left(\beta \sum_{<\alpha,\alpha'>\subset \Lambda} J\sigma_{\alpha}\sigma_{\alpha'}\right)},
\end{equation}
where the sum is over all spin configurations $\sigma$ in $\mathcal{C}_\Lambda=\{\pm 1\}^{\Lambda}$, which satisfy the boundary conditions. For more convenience we set $J=1$.

As we mentioned, the transfer matrix formalism can be used to calculate the partition function and correlation functions of operators such as spin, energy etc. in planar Ising model on the rectangle with the specific boundary conditions. In order to calculate the partition function and correlation functions in transfer matrix formalism, the sums over all configurations in partition function and correlation functions are divided into the multiple sums over the configurations of the rows, $\mathcal{C}_\Lambda\text{(row)}=\{\pm1\}^{2M+1}$.

Let us define the Hilbert space $\mathcal{H}=\bigotimes_{j=-M}^{M}\mathds{C}_j^2$, with basis $\mathbf{e}_\sigma = \bigotimes_{j=-M}^{M}
{\begin{bmatrix}
     \frac{1+\sigma_j}{2}\\
       \frac{1-\sigma_j}{2}
\end{bmatrix}}$, and the spin operator,
\begin{equation} 
 \hat\sigma_j=
 1\otimes...\otimes 1\otimes
\begin{bmatrix}
     1 & 0\\
       0& -1
\end{bmatrix}
 \otimes 1\otimes...\otimes 1=
 {\begin{bmatrix}
      1 & 0\\
       0& -1
\end{bmatrix}}_j,
\end{equation}
acting on $\mathcal{H}$.
The transfer matrix $V_M: \mathcal{H}\rightarrow \mathcal{H}$ is a linear transformation on the Hilbert space, defined by $V_M=V_1^{\frac{1}{2}}V_2V_1^{\frac{1}{2}}$, where $V_1$ and $V_2$ are horizontal and vertical parts of the transfer matrix defined as follows. The action of $V_1$ and $V_2$ on the basis of Hilbert space are defined by
\begin{equation}
V_1 \mathbf{e}_{\sigma}= \exp{\left(\beta\sum_{j=-M}^{M-1}\hat{\sigma}_j\hat{\sigma}_{j+1}\right)}\mathbf{e}_{\sigma},
\hspace{.2cm}
V_2 \mathbf{e}_{\sigma}= \sum_{\rho}V_2\mathbf{e}_{\rho},
\hspace{.2cm}
(V_2)_{\rho\sigma}=e^{-2b} \exp{\left(\sum_{j=-M}^{M}\beta(j)\rho_j\sigma_j\right)},
\end{equation}
where $b\in \mathds{R}$, $\rho, \sigma  \in \mathcal{C}_\Lambda\text{(row)}$, $\rho_j, \sigma_j$ are the row configurations evaluated at $j$-th column, $(V_2)_{\rho\sigma}$ is the matrix element of $V_2$, and
$\beta(j) = \begin{cases}
  b & \text{for } |j| = M \\
  \beta & \text{for } |j| \neq M
\end{cases}$.
We are interested in the action of $V_2$ in the limit $b\to \infty$; $\lim_{b\to \infty}V_2 \mathbf{e}_{\sigma}= \sum_{\rho, \rho_{\pm M}=\sigma_{\pm M}}V_2\mathbf{e}_{\rho}.$

\vspace{.2cm}
It has been obtained in proposition (1.1.1) in \cite{Pal07}, that the partition function (\ref{Ising partition function}), and correlation function of $\mathcal{O}_A= \prod_{i\in A}\mathcal{O}_i$, i.e. a product of linear operators $\mathcal{O}_i$ such as spin etc. in a subset $A$ of the domain $\Lambda$, in the limit $b\to \infty$, are given by
\begin{eqnarray}\label{correlation function definition}
&&Z_{\Lambda}=<\mathbf{e}_{\sigma}^{N}|V_1^{\frac{1}{2}}V_M^{2N}V_1^{\frac{1}{2}}|\mathbf{e}_{\sigma}^{-N}>,\nonumber\\
    &&<\mathcal{O}_A>_\Lambda=\frac{<\mathbf{e}_{\sigma}^{N}|V_1^{\frac{1}{2}}V_M\mathcal{O}_{A_{N-1}}V_M\mathcal{O}_{A_{N-2}}...\mathcal{O}_{A_{-N+1}}V_MV_1^{\frac{1}{2}}|\mathbf{e}_{\sigma}^{-N}>}{Z_{\Lambda}},
\end{eqnarray}
where $\mathcal{O}_{A_i}$ denotes the restriction of $\mathcal{O}_A$ to the $i$-th row, and $\mathbf{e}_{\sigma}^{\pm N}$ is the Hilbert space representation of the $\pm N$-th row configuration. We consider the lower equation in (\ref{correlation function definition}) as a definition of the correlation function.
\subsection{Ising free fermions}
The free fermions were introduced in 40's in order to compute the free energy of the Ising model. This is one of the powerful methods besides other methods such as combinatorial methods, that have led to the integrability paradigm in the Ising model, \cite{Ba08}.

In order to discuss the free fermions in Ising model we introduce a representation of the Clifford algebra in Ising model. Suppose that $W$ is a finite-dimensional complex vector space with a nondegenerate complex bilinear form denoted by $(\cdot, \cdot)$. A Clifford algebra \textit{Cliff}$(W)$ on the vector space $W$ is defined as an associative algebra with unit $e$  and set of generators in $W$ satisfying $ab+ba=(a,b)e$.

We define a complex vector space $W_M'$, with orthonormal basis, Clifford algebra generators $\{\frac{p_k}{\sqrt{2}},\frac{q_k}{\sqrt{2}}\}$, as follows,
\begin{equation}
W_M'=Span(\{p_k|k\in I_M-\frac{1}{2}\}\cup \{q_k|k\in I_M+\frac{1}{2}\})
=W_M\oplus (\mathds{C}_{p_{-M-\frac{1}{2}}}+\mathds{C}_{q_{M+\frac{1}{2}}}),
\end{equation}
where $I_M=\{-M, -M+1, ..., M\}$.
Moreover, we define a finite-dimensional, irreducible spin representation of Clifford algebra \textit{Cliff}$(W_M')$, so called Brauer-Weyl representation, with generators,
\begin{equation}
p_k= \left\{\prod_{j=-M}^{k-\frac{1}{2}}
{\begin{bmatrix}
      0 & 1\\
       1& 0
 \end{bmatrix}}_j
 \right\}
 {\begin{bmatrix}
      1 & 0\\
       0& -1
 \end{bmatrix}}_{k+\frac{1}{2}},\hspace{0.5cm}
 q_k= \left\{\prod_{j=-M}^{k-\frac{3}{2}}
{\begin{bmatrix}
      0 & 1\\
       1& 0
 \end{bmatrix}}_j
 \right\}
 {\begin{bmatrix}
      0 & -\ii\\
       \ii& 0
 \end{bmatrix}}_{k-\frac{1}{2}},
\end{equation}
acting on $\mathcal{H}$. It can be easily checked that $p_k$ and $q_k$ satisfy the anti-commutation relations,
\begin{equation}
\{p_k, p_l\}=2\delta_{kl},\hspace{.2cm}
\{q_k, q_l\}=2\delta_{kl},\hspace{.2cm}
\{p_k, q_l\}=0.
\end{equation}
Then, the lattice fermion $\psi_k$ and anti-fermion $\bar\psi_k$ operators are defined on the mid-points of horizontal edges of two-dimensional rectangular lattice,
\begin{equation}
\psi_k=A_{\psi}(q_k+p_k),\hspace{.5cm}
\bar\psi_k=A_{\bar\psi}(-q_k+p_k),
\end{equation}
where $A_{\psi}$ and $A_{\bar\psi}$ are normalization factors.

The row-to-row propagation/time evolution of the fermions and their correlation functions in the transfer matrix analysis, and their relations to discrete holomorphicity are studied in \cite{HKZ12}.
It has been shown in \cite{HKZ12}, that the transfer matrix can be written in terms of Clifford algebra generators $p_k$ and $q_k$ and thus the time evolution of the free fermions can be explicitly calculated via the conjugation by the transfer matrix. This conjugation is called induced rotation and it is denoted by $T(V_M)$. The induced rotation is a linear transformation, $T(V_M):W_M'\rightarrow W_M'$ such that for all $v\in W_M'$, $T(V_M)v=V_M^{-1}vV_M$. The induced rotation preserves the bilinear form, $(T(V_M)a, T(V_M)b)=(a,b)$ for $a,b \in W$.
Furthermore, the time($m$)-dependent fermion at column $k$ and row $m$ is defined by using the transfer matrix, as $\psi(k+\ii m)=V_M^{-m}\psi_k V_M^m$.
Moreover, Eq. (\ref{correlation function definition}) can be used to define the normalized lattice correlation functions of fermion operators in the domain $\Lambda$ with plus boundary conditions, i.e. plus spins on the boundary of $\Lambda$,
\begin{equation}
    <\prod_{i=1}^{2n}\psi(z_i)>_\Lambda^{(+)}=\frac{1}{Z}<++|\prod_{i=1}^{2n}\psi(z_i)|++>,\hspace{.2cm}\textit{with}\ |++>= V_M^{N}V_1^{\frac{1}{2}}|\mathbf{e}_{(+)}^{-N}>,\ <++|=<\mathbf{e}_{(+)}^{N}|V_1^{\frac{1}{2}}V_M^{N},
\end{equation}
where $z_i=k_i+\ii m_i$, $Z=<++|++>$ is the partition function, and $\mathbf{e}_{(+)}^{\pm N}$ corresponds to a $\pm N$-th row configuration in which all the spins are plus. It has been discussed in sections (1.3) and (4.2) in \cite{Pal07}, that the naive scaling limit of the transfer matrix formalism for the free fermions leads to the Dirac equation and at critical temperature, one can observe that the free fermions (anti-fermions) are holomorphic (anti-holomorphic) functions. In next section, we explain a rigorous approach to derive the scaling limit of the lattice fermion correlation functions.
\subsection{Scaling limit of discrete holomorphic observables and correlation functions}
The methods of discrete complex analysis and discrete holomorphic functions \cite{Smi06}, \cite{Smi10a}, \cite{Smi10b}, \cite{ChSm09} \cite{ChSm11}, \cite{IkCa09} and \cite{RaCa07} provide the possibility to perform the rigorous scaling limit of the s-holomorphic functions and parafermionic observables as well as other advantages. Thus, by using the relations between s-holomorphic functions and fermion correlation functions we can obtain the scaling limit of the correlation functions, rigorously. The obtained results from this method coincide with the vacuum correlation functions of free fermions in conformal field theory. In fact, fermionic CFT describes the continuum limit of the free fermions of the Ising model, \cite{McWu73} and \cite{DMS96}.

In an intuitive sense, the scaling limit of the fermion correlation functions in critical Ising model on a strip with lattice mesh size $\delta$ is defined by taking first the semi-infinite volume limit $N\rightarrow\infty$ and then taking the continuum limit, $M\rightarrow\infty$, $\delta\rightarrow 0$, while the width of the strip $M\delta$ is kept fixed. For example, the scaling limit of boundary state $|++>$ is expected to behave like $|++>\overset{N\rightarrow\infty}{\longrightarrow}|0>_M\overset{M\rightarrow\infty,\delta\rightarrow 0}{\longrightarrow} |0>$, i.e. a CFT vacuum.

The first step towards the rigorous scaling limit of the Ising free fermions on the lattice is to find the scaling limit of the lattice fermion correlation functions. This can be done via the specific functions, called s-holomorphic functions,
\begin{equation}
F_{z'}^{\uparrow}(z)=\frac{1}{\mathcal{Z}}\sum_{\gamma \in\mathcal{C}_{z'^{\uparrow}}} \alpha_c^{L(\gamma)}e^{-\frac{\ii}{2}\mathcal{W}(\gamma:z'\rightsquigarrow z)},\hspace{.3cm}
F_{z'}^{\downarrow}(z)=\frac{1}{\mathcal{Z}}\sum_{\gamma \in\mathcal{C}_{z'^{\downarrow}}} \alpha_c^{L(\gamma)}e^{-\frac{\ii}{2}\mathcal{W}(\gamma:z'\rightsquigarrow z)},
\end{equation}
where $\mathcal{Z}=\sum_{\gamma\in \mathcal{C^+}}\alpha_c^{L(\gamma)}$ is a partition function with plus boundary conditions, and the sum in $F_{z'}^{\uparrow}(z)$ ($F_{z'}^{\downarrow}(z)$) is over collections of dual edges in all graphical expansions $\gamma\in\mathcal{C}_{z'^{\uparrow}}$ ($\gamma\in\mathcal{C}_{z'^{\downarrow}}$) consisting of loops and a path stars at $z'$ upward (downward) and ends at $z$ either from above or below, $\alpha_c=e^{-2\beta_c}$, $L(\gamma)$ is the total number of edges in the configuration $\gamma$ and $\mathcal{W}(z'\rightsquigarrow z)$ is the winding number of directed path starting at $z'$ and ending at $z$. The points $z,z'$ are midpoints of horizontal edges of the lattice. These functions are called parafermionic observables of the Ising model. They are solutions of the Riemann boundary value problem. For a general definition of Riemann boundary value problem see section (2) of \cite{HKZ12}. Then we need to obtain the scaling limit of the above s-holomorphic functions.
  
It has been shown that the scaling limit of the s-holomorphic functions which satisfies the Riemann boundary conditions exists and the convergence of the parafermionic observables as $\delta\rightarrow 0$ can be controlled by the methods of discrete complex and harmonic analysis, \cite{HoSm10b, Hon10a}. The result of these studies can be summarized as follows, the functions $\frac{F_{z'}^{\uparrow}(z)}{\delta}, \frac{F_{z'}^{\downarrow}(z)}{\delta}$ converge uniformly on compact subsets of $D\setminus\{z'\}$ to the unique holomorphic functions, $\lim_{\delta\rightarrow0}\frac{F_{z'}^{\uparrow}(z)}{\delta}= f_{z'}^{\uparrow}(z)$ and $\lim_{\delta\rightarrow0}\frac{F_{z'}^{\downarrow}(z)}{\delta}= f_{z'}^{\downarrow}(z)$. These functions satisfy the following continuum Riemann boundary value problem,
\begin{equation}\label{RHP}
    \begin{cases}
  f_{z'}^{\uparrow}(z)\  \textit{and} \ f_{z'}^{\downarrow}(z) \ \textit{are holomorphic on rectangle}\ \setminus \{z'\} \\
  2\pi \ii\, \mathrm{Res}_{z=z'} f_{z'}^{\uparrow}(z)= -1, \ 2\pi \ii\, \mathrm{Res}_{z=z'} f_{z'}^{\downarrow}(z)= 1  \\
  \textit{for}\ z\in \partial_{\textit{rectangle}}, \ f_{z'}^{\uparrow}(z) \parallel \frac{1}{\sqrt{-\nu_z}}, \ f_{z'}^{\downarrow}(z) \parallel \frac{1}{\sqrt{-\nu_z}}
\end{cases},
\end{equation}
where $\partial_{\textit{rectangle}}$ is the boundary of the rectangle and $\nu_z$ is the counter clock-wise tangent vector at point $z$ on the boundary of the rectangle. To obtain the above statement, the Riemann boundary conditions are considered and the residue calculations on the lattice are performed by considering couple of combinatorial cases and using the fact that the contour integral of s-holomorphic function is zero.
This Riemann boundary value problem determines a unique function on the rectangle which transforms conformally covariant under the conformal transformations between the rectangle and any other domains.
The holomorphic functions on the half-plane which satisfy (\ref{RHP}) can be obtained as
\begin{equation}\label{holomorphic functions half plane}
    f_{z'}^{\uparrow;\mathbb{H}}(z)= \frac{\ii}{2\pi}\left(\frac{1}{z-z'}+\frac{1}{z-\bar{z}'} \right),\hspace{.3cm}
    f_{z'}^{\downarrow;\mathbb{H}}(z)= \frac{\ii}{2\pi}\left(\frac{-1}{z-z'}+\frac{1}{z-\bar{z}'} \right).
\end{equation}

The scaling limit of the correlation functions of Ising fermions on the upper-half plane can be determined by using first, the relation between lattice fermionic correlation functions and parafermionic observables and second, the scaling limit of the parafermionic observables.
The relations between fermionic correlation functions and parafermionic observables are stated in Theorem (22) in \cite{HKZ12}. In  slightly different notation they are as follows,
\begin{eqnarray}
    <++|\psi(z)\psi(z')|++>&=&2 A_\psi^2Z(F_{z'}^{\uparrow}(z)-F_{z'}^{\downarrow}(z)),\nonumber\\
    <++|\psi(z)\bar\psi(\bar {z'})|++>&=&2\ii A_\psi A_{\bar\psi}Z(F_{z'}^{\uparrow}(z)+F_{z'}^{\downarrow}(z)),\nonumber\\
    <++|\bar\psi(\bar z)\bar\psi(\bar {z'})|++>&=&2 A_{\bar\psi}^2Z(\overline{F_{z'}^{\uparrow}(z)}-
    \overline{F_{z'}^{\downarrow}(z)}).
\end{eqnarray}

By using the above relations and the convergence of the scaling limit of the parafermionic observables, Eq. (\ref{holomorphic functions half plane}), one can obtain the scaling limit of the lattice fermion correlation functions,\newline
$\lim_{\delta\rightarrow0}\frac{1}{\delta}<++|\psi(z)\psi(z')|++>$, by transformation from the strip to the half-plane, as follows,
\begin{equation}
<\psi(z)\psi(z')>_{\mathbb{H}}=2A_\psi^2Z(f_{z'}^{\uparrow;\mathbb{H}}(z)-f_{z'}^{\downarrow;\mathbb{H}}(z)).
\end{equation}
With the choice of parameters, $A_\psi=\frac{1}{\sqrt{2}}(-\ii-1), A_{\bar \psi}=\overline{A_\psi}$ and $Z=-\frac{\pi}{2}$, we have,
\begin{equation}\label{two point corr func half plane}
     <\psi(z)\psi(z')>_{\mathbb{H}}= \left(\frac{1}{z-z'}\right),\hspace{.2cm}
     <\psi(z)\bar\psi(\bar {z'})>_{\mathbb{H}}=\left(\frac{1}{z-\bar{z'}}\right),\hspace{.2cm}
     <\bar\psi(\bar z)\bar\psi(\bar {z'})>_{\mathbb{H}}=\left(\frac{1}{\bar z-\bar{z'}}\right).
\end{equation}
So far we have discussed only the rectangular domain but the parafermionic observables can be defined similarly in any square lattice domain \cite{HoSm10b}. However, we studied the scaling limit of the parafermionic observables on the rectangle at the critical point, $\beta = \beta_c$. In general, a continuous domain $D$ can be approximated with the discrete domain $D_\delta$, as a subgraph of the square lattice $\delta \mathds{Z}^2$, when the small lattice mesh size $\delta$ tends to zero, $\delta\rightarrow 0$.

To summarize, we observed that the scaling limit of the two-point correlation function of the Ising free fermions is equal to the two-point correlation function of the fermionic CFT. Thus, in the scaling limit, it would be natural to think of the Ising fermions as fermions of CFT.
\section{Fermionic conformal field theory on domain D}
In this part we study a two-dimensional fermionic boundary conformal field theory on domain $D$, i.e. a field theory which describes the scaling limit of the free fermions of Ising model on domains with boundaries.
The goal of this part is to construct a fermionic Fock space of fields and to study their properties, such as behavior of the correlation functions, operator product expansions and differential equations, which will be used in the CFT/SLE correspondence for the Ising model. We study the holomorphic part of the theory and the anti-holomorphic part is similar.
\subsubsection*{Conformal transformations}
In field theories which are relevant for studies about statistical mechanics, domain conformal transformation $h:D\rightarrow D'$ is considered. Roughly speaking, the values of the fields on domain $D$, $\psi(z)$ for $z\in D$ conformally transform to values of the fields on domain $D'$, $\psi(h(z))$ for $h(z)\in D'$. Fermions in CFT are defined by their transformation rule. By definition, the fermion field $\psi(z)$ is a conformal primary field of dimension $1/2$ that satisfies
\begin{equation}
  \psi(z) = h'(z)^{\frac{1}{2}}\psi(h(z)),
\end{equation}
where $h'(z)$ is derivative of $h(z)$ with respect to $z$. We will return to conformal transformation of the fermions when we will discuss the correlation functions on domain $D$.
\subsubsection*{Fermionic Fock space of fields $\mathcal{F}$}
We defined the scaling limit of the Ising fermions, the free fermion fields $\psi(z)$. Other fermionic fields, called descendant fields, are Fock space fields in domain $D \subset \mathds{C}$ and they are constructed by normally ordered product of derivatives of free fermion fields, e.g. $:\partial^2\psi(z)\partial\psi(z)\psi(z):$.

On the half-plane, a finite Fock space field $X_k(z)= :\partial^{k_n}\psi(z)\partial^{k_{n-1}}\psi(z)...\partial^{k_2}\psi(z)\partial^{k_1}\psi(z):\in \mathcal{F}$, for $z\in \mathbb{H}$, is defined by
\newpage
\begin{equation}\label{def Fock space field half plane}
   :\partial^{k_n}\psi(z)\partial^{k_{n-1}}\psi(z)...\partial^{k_2}\psi(z)\partial^{k_1}\psi(z):= \partial^{k_n}\psi(z)\partial^{k_{n-1}}\psi(z)...\partial^{k_2}\psi(z)\partial^{k_1}\psi(z)\nonumber
\end{equation}
\begin{eqnarray}
    -\lim_{z_{i_m}, z_{j_m}\to z}\sum_{s=1}^{n}\sum_{i_1<...<i_s, j_1\neq...\neq j_s}
    (&&\pm\partial^{k_{i_1}}\partial^{k_{j_1}}[\frac{1}{z_{i_1}-z_{j_1}}]...
    \partial^{k_{i_s}}\partial^{k_{j_s}}[\frac{1}{z_{i_s}-z_{j_s}}]\nonumber\\
    &&:\partial^{k_n}\psi(z)\partial^{k_{n-1}}\psi(z)...\partial^{k_2}\psi(z)\partial^{k_1}\psi(z):_{(i_1,...,i_s;j_1,..., j_s)}),
\end{eqnarray}
where $m=1, ..., s$, and the order of the limits is from right to left, $\partial^{k_i}=\frac{1}{(k_i+1)!}\frac{\partial^{k_i}}{\partial z^{k_i}}$, and the subscript $(i_1,...,i_s;j_1,..., j_s)$ means that the fields $\partial^{k_{i_1}}\psi(z), ..., \partial^{k_{i_s}}\psi(z)$ and $\partial^{k_{j_1}}\psi(w), ..., \partial^{k_{j_s}}\psi(w)$ are removed. We used the two-point correlation function of fermion on half-plane, Eq. (\ref{two point corr func half plane}) in the above definition. However, definition of a general Fock space field on domain $D$ is more complicated and requires removing all the divergent parts on domain $D$. The transformation rules for general conformal Fock space fields are more complicated than that of free fermion field,
\begin{equation}
  X(z) = h'(z)^{\lambda_X}X(h(z))+...,
\end{equation}
where $\lambda_X$ is the conformal dimension of $X$ and ... represents complicated function of higher order derivatives of $h(z)$. We will see in section (3.1), an example of the transformation rule for a special field, called Virasoro field.

The multiplication of the two general Fock space fields on half-plane can be obtained by using the Wick's theorem, \cite{DMS96}, as follows,
\begin{equation}\label{Wicks formula half plane}
    :\partial^{k_n}\psi(z)\partial^{k_{n-1}}\psi(z)...\partial^{k_2}\psi(z)\partial^{k_1}\psi(z):
    :\partial^{l_m}\psi(w)\partial^{l_{m-1}}\psi(w)...\partial^{l_2}\psi(w)\partial^{l_1}\psi(w):=\nonumber
\end{equation}
\begin{equation}
    \sum_{s=0}^{min(n,m)}\sum_{i_1<...<i_s, j_1\neq...\neq j_s}
    (\pm <\partial_z^{k_{i_1}}\psi(z)\partial_w^{l_{j_1}}\psi(w)>_{\mathbb{H}}...
    <\partial_z^{k_{i_s}}\psi(z)\partial_w^{l_{j_s}}\psi(w)>_{\mathbb{H}}\nonumber
\end{equation}
\begin{equation}
    :\partial^{k_n}\psi(z)\partial^{k_{n-1}}\psi(z)...\partial^{k_2}\psi(z)\partial^{k_1}\psi(z)
    \partial^{l_m}\psi(w)\partial^{l_{m-1}}\psi(w)...\partial^{l_2}\psi(w)\partial^{l_1}\psi(w):_{(i_1,...,i_s;j_1,..., j_s)}),
\end{equation}
where the contraction in the Wick's formula is given by $<\partial_z^{k_{i_s}}\psi(z)\partial_w^{l_{j_s}}\psi(w)>_{\mathbb{H}}=\partial_z^{k_{i_s}}\partial_w^{l_{j_s}}[\frac{1}{z-w}]$ for $z, w\in\mathbb{H}$.
\subsubsection{Correlation functions}
Using the discrete holomorphicity results, the two-point correlation functions of free fermion fields $\psi(z)$, on the half-plane, Eq. (\ref{two point corr func half plane}), are obtained rigorously in section (2.3). In the scaling limit on the upper-half plane, by using the two-point correlation functions of fermions $\psi(z)$, any $2n$-point correlation function can be written in terms of two-point functions, via the Wick's theorem,
\begin{equation}
    <\psi(z_1)...\psi(z_{2n})>_{\mathbb{H}}=Pf((<\psi(z_i)\psi(z_j)>_{\mathbb{H}})_{i,j=1}^{2n})= Pf\left(\left[\frac{1}{z_i-z_j}\right]_{i,j=1}^{2n}\right),
\end{equation}
where Pf is the Pfaffian.
The Pfaffian of an anti-symmetric matrix $A\in \mathds{C}^{n\times n}$ is defined by
\begin{equation}
Pf(A)=
\begin{cases}
  \frac{1}{2^kk!}\sum_{P} Sgn(P) \prod_{i=1}^{k}A_{P(2i-1), P(2i)} & \text{for } n =2k \\
  0 & \text{for } n=2k-1
\end{cases},\nonumber
\end{equation}
where $P$ is any permutation of $\{1,2,...,2n\}$ and $Sgn(P)$ is the sign of the permutation.
This result can be proved by using the Pfaffian formula for the lattice fermion correlation function (section (4.4) in \cite{HKZ12}) and then taking the scaling limit.

We can obtain the two-point correlation functions in an arbitrary domain $D$ by using the fermion transformation rule $ \psi(z) = g'(z)^{\frac{1}{2}}\psi(g(z))$, under a conformal map $g: D\rightarrow \mathbb{H}$,
\begin{equation}
      <\psi(z)\psi(z')>_{D}= \frac{g'(z)^{\frac{1}{2}}g'(z')^{\frac{1}{2}}}{g(z)-g(z')},\hspace{.2cm}
      <\psi(z)\bar\psi(\bar {z'})>_D= \frac{g'(z)^{\frac{1}{2}}\overline{g'(z')}^{\frac{1}{2}}}{g(z)-\overline{g(z')}},\hspace{.2cm}
      <\bar\psi(\bar z)\bar\psi(\bar {z'})>_D=\frac{\overline{g'(z)}^{\frac{1}{2}}\overline{g'(z')}^{\frac{1}{2}}}{\overline{g(z)}-\overline{g(z')}}.
\end{equation}
In the following, we find asymptotic results for the correlation functions of fermions on domain $D$ by using the Laurent expansion of the function $g(z):D\rightarrow \mathbb{H}$ and its derivative up to some fixed order. Up to a fixed order, one can check that
\begin{equation}
    [g(z)-g(w)]^{-1}= \frac{1}{\epsilon g'(w)} (1-\frac{\epsilon g''(w)}{2g'(w)}-\frac{\epsilon^2 g'''(w)}{6g'(w)}+\frac{\epsilon^2 g''^2(w)}{4g'^2(w)}),\nonumber
\end{equation}
\begin{equation}
    g'(z)^{\frac{1}{2}}g'(w)^{\frac{1}{2}}= g'(w) (1+\frac{\epsilon g''(w)}{2g'(w)}+ \frac{\epsilon^2g'''(w)}{4g'(w)}-\frac{\epsilon^2g''^2(w)}{8g'^2(w)}),
\end{equation}
where $\epsilon= z-w$.
These expansions lead to an asymptotic formula for the two-point function of fermions in the domain $D$,
\begin{equation}
    <\psi(z)\psi(w)>_{D}=\frac{\sqrt{g'(z)}\sqrt{g'(w)}}{g(z)-g(w)}=
    \frac{1}{z-w}+(z-w)\left(\frac{1}{12}\frac{g'''(w)}{g'(w)}-\frac{1}{8}(\frac{g''(w)}{g'(w)})^2\right)+...\nonumber
\end{equation}
\begin{equation}\label{two point correlation in D}
    =<\psi(z)\psi(w)>_{\mathbb{H}}+\frac{(z-w)}{12}S_g(w)+...,
\end{equation}
where $S_g(w)=\frac{g'''(w)}{g'(w)}-\frac{3}{2}(\frac{g''(w)}{g'(w)})^2$ is the Schwarzian derivative of function $g$.
By using the above results, the $2n$-point correlation functions of free fermion fields $\psi(z)$ are obtained,
\begin{equation}
    <\prod_{i=1}^{2n}\psi(z_i)>_D = Pf((<\psi(z_i)\psi(z_j)>_D)_{i,j=1}^{2n})= Pf\left(\left[\frac{\sqrt{g'(z_i)}\sqrt{g'(z_j)}}{g(z_i)-g(z_j)}\right]_{i,j=1}^{2n}\right).
\end{equation}
By using the Pfaffian formula for the higher point correlation functions, we can observe that,
\begin{equation}
    <\psi(z_1)...\psi(z_m)>_{D}\sim <\psi(z_1)...\psi(z_m)>_{\mathbb{H}},
\end{equation}
only holds for $m\leq 4$ and $\sim$ means that the two sides have the same divergent terms in the limit $z\to w$.

All the other correlation functions of fermionic Fock space fields, $<X_1(z_1)...X_n(z_n)>_D$, can be obtained from the correlation functions of free fermion fields $\psi(z)$ by using the Wick's theorem and taking the derivatives of the two-point fermion correlation functions.
The correlation function of derivatives of fermion fields is simply given by
\begin{equation}
    <(\frac{\partial^{m_1}}{\partial z_1^{m_1}})\psi(z_1)...(\frac{\partial^{m_{2n}}}{\partial z_{2n}^{m_{2n}}})\psi(z_{2n})>_D= \frac{\partial^{m_1+...+m_{2n}}}{\partial z_1^{m_1}...\partial z_{2n}^{m_{2n}}}Pf\left(\left[\frac{\sqrt{g'(z_i)}\sqrt{g'(z_j)}}{g(z_i)-g(z_j)}\right]_{i,j=1}^{2n}\right).
\end{equation}
\subsubsection*{Operator product expansions}
The OPE between two Fock space fields is an expansion of the Wick's formula on domain $D$, (see lectures (1) and (2) in \cite{KaMa11}), when the positions of two fields become close. Notice that, in general the OPE is domain dependent. Thus, OPE is an asymptotic expansion of $X(z)Y(w)$ on domain $D$ as $z\rightarrow w$,
\begin{equation}
    X(z)Y(w)= \sum_{n\in \mathbb{Z}} C_n(w) (z-w)^n ;\  \textit{as}\ z\rightarrow w,
\end{equation}
where the OPE coefficients $C_n$, denoted by $X*_nY$, are also Fock space fields, for further description see lecture (2) in \cite{KaMa11}. We define an OPE product as $X*Y$ where $*_0=*$. Moreover, the singular part of the OPE is defined by
\begin{equation}
    X(z)Y(w)\sim \sum_{n<0} C_n(w) (z-w)^n.
\end{equation}

In the case of fermionic CFT, the formal OPE between free fermions is
\begin{equation}
    \psi(z)\psi(w)= \sum_{n\in \mathbb{Z}} c_n(w)(z-w)^n,
\end{equation}
and this can be explicitly determined by means of Wick's formula on domain $D$,
\begin{equation}\label{fermion Wicks formula on D}
    \psi(z)\psi(w)= <\psi(z)\psi(w)>_D+\psi(z)\odot\psi(w)= \frac{1}{z-w}+\psi(z)\odot\psi(w)+ reg(D),
\end{equation}
where $\psi(z)\odot\psi(w)$ is called normal order product in domain $D$ and $reg(D)$ denotes the terms which do not diverge in the limit $z\rightarrow w$.
Thus, the singular part of the OPE is given by the first term since the other terms vanish as $z \rightarrow w$ and thus we have $\psi(z)\psi(w)\sim \frac{1}{z-w}$.

\subsection{Virasoro algebra representation of the CFT}
The underlying algebraic structure of the CFT is Virasoro algebra. In fact, fermionic conformal field theory is a representation of the Virasoro algebra and Clifford algebra. In this part we review the Virasoro algebra and its representation for the Virasoro generators and Virasoro fields. For further descriptions see lecture (5) in \cite{KaMa11}.

As a special example of the general definition of Fock space fields we define the fermionic Virasoro field. By using the definition of OPE product we define the Virasoro field, $T(z)= -\frac{1}{2}\psi(z)*\partial \psi(z)$. Thus, on the half-plane, the Virasoro field is
\begin{equation}
    T(z)= -\frac{1}{2}:\psi(z)\partial_z\psi(z):=\lim_{w\rightarrow z}[-\frac{1}{2}(\psi(z)\partial_w\psi(w)-\partial_w(\frac{1}{z-w}))].
\end{equation}
And on the domain $D$,
\begin{eqnarray}\label{Virasoro field on D}
    T(z)= -\frac{1}{2}\psi(z)\odot\partial_z\psi(z)&=&\lim_{w\rightarrow z}[-\frac{1}{2}(\psi(z)\partial_w\psi(w)-\partial_w(<\psi(z)\partial_w\psi(w)>_D))]\nonumber\\
    &=&\lim_{w\rightarrow z}[-\frac{1}{2}(\psi(z)\partial_w\psi(w)-\partial_w(\frac{1}{z-w}))]-\frac{1}{24} S_g(z),
\end{eqnarray}
where we used Eq. (\ref{two point correlation in D}) to obtain the second line. Thus, the Virasoro field has the same divergent part on the half-plane and domain $D$.

Transformation rule for Virasoro field can be obtained by direct computation in Taylor expansion of the $h(z)$, starting from the fermion transformation rule and definition of the Virasoro field, as follows,
\begin{equation}
     T(z) =  h'(z)^{2}T(h(z))+ \frac{1}{24}S_h(z).
\end{equation}
The Virasoro field is called conformal quasi-primary field since it satisfies the above equation.

By using the Wick's formula and Taylor expansion, direct computations shows that the Virasoro field in a CFT with central charge $c$, satisfies the Virasoro operator product expansion on the half-plane,
\begin{equation}\label{OPE Virasoro field on half plane}
    T(z)T(w)\sim\frac{c/2}{(z-w)^4}+ \frac{2}{(z-w)^2}T(w)+\frac{1}{(z-w)}\partial_wT(w).
\end{equation}

The Virasoro operator is defined by,
\begin{equation}
    L_n(z)= \frac{1}{2\pi \ii}\oint_{(z)}(\zeta-z)^{1+n}T(\zeta)d\zeta.
\end{equation}
One can check that the Virasoro field OPE, Eq. (\ref{OPE Virasoro field on half plane}) leads to the Virasoro algebra,
\begin{equation}
    [L_m,L_n]=(m-n)L_{m+n}+\frac{c}{12}m(m^2-1)\delta_{m+n,0},
\end{equation}
where $c=12\mu$ and $\mu$ is the order of $T$ as a Schwarzian form, which will be defined in the following.

In order to obtain a Virasoro algebra representation in the space of all Fock space fields in domain $D$, the action of Virasoro operator $L_n$ on Fock space fields such as $X$ is defined as,
\begin{equation}
    L_nX=T*_{(-n-2)}X.
\end{equation}
At the end of this section we describe the OPE between Virasoro field and primary fields. From the explicit form of the OPE of Virasoro field $T$ and a primary field $X$, one can check that for $n\geq -1$, \cite{KaMa11},
\begin{equation}
    (L_n X)(z)=
    (v_n\partial_z+\lambda v'_n)X(z),
\end{equation}
where $v_n(z)= (\zeta-z)^{1+n}$, $v'_n$ is the derivative of $v_n$ with respect to $z$ and $\lambda$ is the conformal dimension of $X$.
And for $n\leq -2$ we have,
\begin{equation}
    L_n= \frac{\partial^{-n-2}T}{(-n-2)!}*.
\end{equation}
A Fock space field $X$ is called a primary field with the conformal dimension $\lambda$ if it satisfies,
\begin{equation}\label{Virasoro modes acting on fields}
    L_nX=0, \hspace{.5cm} L_0X=\lambda X, \hspace{.5cm} L_{-1}X=\partial X,
\end{equation}
for $n\geq 1$. For example, the free fermion field $\psi$ is a primary field of conformal dimension $1/2$ in CFT with central charge $c=1/2$. Furthermore, a Schwarzian form $Y$ of order $\mu$ in CFT with $c=12\mu$ is defined by,
\begin{equation}
    L_mY=0, \hspace{.3cm}L_2Y=6\mu I,\hspace{.3cm} L_1Y=0,\hspace{.3cm}  L_0Y=2 Y, \hspace{.3cm} L_{-1}Y=\partial Y,
\end{equation}
for $m\geq 3$. For example, the fermionic Virasoro field $T$ is a Schwarzian form of order $1/24$ in CFT with central charge $c=1/2$ and it is called quasi-primary field.

In a conformal field theory with central charge $c= \frac{2\lambda}{2\lambda+1}(5-8\lambda)$ and a primary field $X$ with conformal dimension $\lambda$, the Fock space field $X_s= (L_{-2}-\frac{3}{2(2\lambda+1)}L_{-1}^2)X$ is also a primary field and is called singular vector at level two of the Virasoro algebra representation. The field $X$ is called a primary field degenerate (non-degenerate) at level two if $X_s=0$ ($X_s\neq 0$). Moreover, $L_0X_s=(\lambda+2)X_s$.

Free fermionic field theory of fermionic Fock space fields as a conformal field theory is a representation of the Virasoro algebra and Clifford algebra. Straightforward computation shows that in fermionic CFT with $c=1/2$, the fermion field $\psi(z)$ with conformal dimension $h=\frac{1}{2}$ is a primary field degenerate at level two,
\begin{equation}\label{singular field}
    (L_{-2}-\frac{3}{4}L_{-1}^2)\psi(z)=0.
\end{equation}

\subsubsection*{Ward identity and null field differential equation}
In this part, we continue our review on the standard results in CFT, such as Ward identity and null field differential equation.
The Ward identity in $\mathbb{H}$ for the correlation functions of Gaussian free fields is obtained in \cite{KaMa11}. Similarly, for the fermionic fields we have,
\begin{equation}\label{Ward indentity on half plane}
    <T(z)\psi(w_1)...\psi(w_n)>_{\mathbb{H}}= \sum_{i=1}^N\left[\frac{1/2}{(z-w_i)^2}+\frac{1}{z-w_i}\frac{\partial}{\partial w_i}\right]<\psi(w_1)...\psi(w_n)>_{\mathbb{H}}.
\end{equation}
Furthermore, by using $T(z)=g'(z)^2T(g(z))+\frac{1}{24}S_g(z)$ and $\psi(z)=g'(z)^{1/2}\psi(g(z))$, one can obtain a Ward-like identity on domain $D$,
\begin{equation}
    <T(z)\psi(w_1)...\psi(w_n)>_{D}=\nonumber
\end{equation}
\begin{equation}
     \left(\sum_{i=1}^N\left[\frac{g'(z)^2}{2(g(z)-g(w_i))^2}+\frac{g'(z)^2g'(w_i)^{-1}}{g(z)-g(w_i)}\frac{\partial}{\partial w_i}\right]+\frac{1}{24}S_g(z)\right)<\psi(w_1)...\psi(w_n)>_{D}.\nonumber
\end{equation}
Similar to the computation of correlation functions on domain $D$, by using,
\begin{equation}
    [g(z)-g(w)]^{-2}= \frac{1}{\epsilon^2 g'^2(w)} (1-\frac{\epsilon g''(w)}{g'(w)}-\frac{\epsilon^2 g'''(w)}{3g'(w)}+\frac{3\epsilon^2 g''^2(w)}{4g'^2(w)}),\nonumber
\end{equation}
\begin{equation}
  [g(z)-g(w)]^{-1}= \frac{1}{\epsilon g'(w)} (1-\frac{\epsilon g''(w)}{2g'(w)}-\frac{\epsilon^2 g'''(w)}{6g'(w)}+\frac{\epsilon^2 g''^2(w)}{4g'^2(w)}),\nonumber
\end{equation}
\begin{equation}
    g'(z)^2= g'(w)^2 (1+2\frac{\epsilon g''(w)}{g'(w)}+ \frac{\epsilon^2g'''(w)}{g'(w)}+\frac{\epsilon^2g''^2(w)}{g'^2(w)}),
\end{equation}
one can obtain,
\newpage
\begin{equation}
   <T(z)\psi(w_1)...\psi(w_n)>_{D}=
\end{equation}
\begin{eqnarray}
   &&(\sum_{i=1}^N\left[\frac{1/2}{(z-w_i)^2}+ \frac{1}{z-w_i}(\frac{1}{2}\frac{g''}{g'}+\frac{\partial}{\partial w_i})+(z-w_i)(S_g(w)-\frac{1}{6}\frac{g'''}{g'}+\frac{7}{4}(\frac{g''}{g'})^2)\frac{\partial}{\partial w_i}+\frac{3}{2}\frac{g''}{g'}\frac{\partial}{\partial w_i}\right]\nonumber\\
&&+(\frac{1}{8}S_g(w)+\frac{1}{4}\frac{g'''}{g'}))<\psi(w_1)...\psi(w_n)>_{D}.\nonumber\\
\end{eqnarray}

It is known that all the correlation functions of descendant Fock space fields can be reduced to correlation functions of primary Fock space fields through the action of differential operators, \cite{DMS96},
\begin{equation}\label{generalized Ward identity}
    <\psi^{(-k_n, -k_{n-1}...-k_2, -k_1)}(z)\prod_{i=1}^N\psi(w_i)>_{\mathbb{H}}=\mathcal{L}_{-k_n}...\mathcal{L}_{-k_1}<\psi(z)\prod_{i=1}^N\psi(w_i)>_{\mathbb{H}},
\end{equation}
where $\psi^{(-k_n, -k_{n-1}...-k_2, -k_1)}(z)=L_{-k_n}L_{-k_{n-1}}...L_{-k_{1}}\psi(z)= \frac{1}{2\pi\ii}\oint dw \frac{1}{(w-z)^{n-1}}T(w)(L_{-k_{n-1}}...L_{-k_1}\psi(z))$ and $\mathcal{L}_{-n}$ is defined by,
\begin{equation}
    \mathcal{L}_{-n}= \sum_{i=1}^N \left\{\frac{1}{2} \frac{(n-1)}{(w_i-z)^n}-\frac{1}{(w_i-z)^{n-1}}\partial_{w_i}\right\}.
\end{equation}
The correlation functions including more than one descendant fields can be explicitly calculated similarly. This means that any correlation function of descendant fields can be reduced to the correlation functions of primary fields. We will use this observation later.

Using Eq. (\ref{singular field}), one can insert the relation $L_{-2}\psi(z)=\frac{3}{4}L_{-1}^2\psi(z)=T*\psi(z)$ in the Ward identity (\ref{Ward indentity on half plane}) to obtain the null field differential equation on the half-plane $\mathbb{H}$,
\begin{equation}\label{null field differential equation}
   \left[\frac{3}{4}\frac{\partial^2}{\partial z^2}-\sum_{i=1}^N\left(\frac{1/2}{(z-w_i)^2}+\frac{1}{z-w_i}\frac{\partial}{\partial w_i}\right)\right] <\psi(z)\psi(w_1)...\psi(w_n)>_{\mathbb{H}}=0.
\end{equation}

Since fermion field and Virasoro field have the same divergent part on half-plane and domain $D$, we expect that the OPE on the domain $D$ is the same as OPE on half-plane. Moreover, by simple calculations using the definition of the Virasoro field Eq. (\ref{Virasoro field on D}), Wick's formula on domain $D$, Eq. (\ref{fermion Wicks formula on D}), and Taylor expansion, the OPE of fermion fields and Virasoro fields for $z,w \in D$ can be obtained,
\begin{equation}\label{OPE on domain I}
    \psi(z)\psi(w)\bigg\vert_D= \frac{1}{z-w}+reg(D),
\end{equation}
\begin{equation}\label{OPE on domain II}
    T(z)\psi(w)\bigg\vert_D= \frac{1/2}{(z-w)^2}\psi(w)+\frac{1}{z-w}\partial_w\psi(w)+\frac{3}{4}\partial_w^2\psi(w)+reg(D),
\end{equation}
\begin{equation}\label{OPE on domain III}
    T(z)T(w)\bigg\vert_D=\frac{1/4}{(z-w)^4}+ \frac{2}{(z-w)^2}T(w)+\frac{1}{(z-w)}\partial_wT(w)+reg(D).
\end{equation}
Thus, by comparing the above results and the known CFT results about the OPE on the half plane we observe that the OPE singular parts of fermion and Virasoro fields are domain independent,
\begin{equation}
    \psi(z)\psi(z')\bigg\vert_D\sim \psi(z)\psi(z')\bigg\vert_{\mathbb{H}}, \hspace{.2cm}T(z)\psi(z')\bigg\vert_D\sim T(z)\psi(z')\bigg\vert_{\mathbb{H}}, \hspace{.2cm}T(z)T(z')\bigg\vert_D\sim T(z)T(z')\bigg\vert_{\mathbb{H}}.
\end{equation}

Consequently, by using the above OPE results, the singular parts of the correlation functions of an arbitrary operator $\mathcal{O}$, fermion fields and Virasoro fields on domain $D$ are given by,
\begin{equation}
    <\psi(z)\psi(w)\mathcal{O}>_D= \frac{<\mathcal{O}>_D}{(z-w)}+ reg(D),
\end{equation}
\begin{equation}
    <T(z)\psi(w)\mathcal{O}>_D= \frac{1/2<\psi(w)\mathcal{O}>_D}{(z-w)^2}+\frac{<\partial_w\psi(w)\mathcal{O}>_D}{z-w}+\frac{3}{4}<\partial_w^2\psi(w)\mathcal{O}>_D+reg(D),
\end{equation}
\begin{equation}
    <T(z)T(w)\mathcal{O}>_D=\frac{1/4}{(z-w)^4}+ \frac{2<T(w)\mathcal{O}>_D}{(z-w)^2}+\frac{<\partial_wT(w)\mathcal{O}>_D}{(z-w)}+reg(D).
\end{equation}
\section{Fermionic vertex operator algebra and its relation to CFT}
In this section, the vertex operator algebra and conformal field theory of the Ising free fermions are studied. The Fock space of fermionic states and fields are constructed in explicit forms and furthermore, the relation between them is investigated.
\subsection{Clifford vertex operator algebra}
In this part, the basic definitions of fermionic vertex operator algebra (FVOA) which is a rigorous algebraic approach to fermionic CFT is reviewed. The FVOA was introduced by R. Borcherds in order to provide a rigorous mathematical definition of the chiral algebra, the symmetry of the two-dimensional CFT and its features such as operator product expansion \cite{Bo86}.
We start with definition of axioms of the fermionic conformal vertex operator algebra, i.e. a conformal VOA which has Clifford algebra symmetry \cite{Ka98}, \cite{Ga06} and \cite{Scht08}.
The Clifford VOA leads to the Fock space of fermionic states and their corresponding fields in the Ising model.

The Fock space of fermionic states is defined as a graded vector space $V=\bigoplus_{n=-\infty}^{\infty}V_n$ consisting of vacuum state $\textbf{1}\in V$ and other states which are generated from the vacuum state and we denote them by small letter such as $a,b,...\in V$.

\subsubsection*{Fermionic vertex operator algebra}
A quadruple $(V, Y, \partial, \textbf{1})$ is called vertex algebra if for all $a\in V$ there exists a mapping $Y: V \rightarrow End(V)\left[[z,z^{-1}]\right]$, $Y(a,z)= \sum_{n\in \mathbb{Z}+\frac{1}{2}}a_n z^{-n-\frac{1}{2}}:=a(z)$ satisfying the following axioms:

$\bullet$ \textsl{Vacuum}: $Y(\textbf{1},z)=I_V$ is the identity;

$\bullet$ \textsl{State-operator correspondence}:
 \begin{equation}
 Y(a,z)\textbf{1}|_{z=0}=a, \Rightarrow \ a_{n}\textbf{1}=0\ for\ n\geq 0\ and\ a_{-\frac{1}{2}}\textbf{1}=a;
 \end{equation}

$\bullet$ \textsl{Translation}:
\begin{equation}
[T, Y(a,z)]= \partial_z Y(a,z),\hspace{.3cm} [T, a_{n}]= (-n+\frac{1}{2})a_{n-1};
\end{equation}
where $T \in End(V)$ is defined by $T(a)= a_{-\frac{3}{2}}\textbf{1}$.

$\bullet$ \textsl{Locality}: $(z-w)^N\{Y(a,z),Y(b,w)\}=0$ for some large $N$;

$\bullet$ \textsl{Regularity}: There is an $M$ such that $a_{n}b=0$, for all $n\geq M$;

In the following we review some results in VOA from \cite{Ka98}, which will be used later.
\subsubsection*{Normal order product in VOA}\cite{Ka98} (Theorem 2.3)
Let us introduce the following notations,
\begin{equation}
    a(z)_-=\sum_{n>0}a_{n}z^{-n-\frac{1}{2}},\hspace{.5cm} a(z)_+=\sum_{n<0}a_{n}z^{-n-\frac{1}{2}}.
\end{equation}
Then, the normal order product of two fields is defined by,
\begin{equation}
    :a(z)b(w):= a(z)_+b(w)-b(w)a(z)_-=a(z)b(w)-\{a(z)_-,b(w)\}.
\end{equation}
The normal order product of more than two fields is defined inductively from right to left as follows,\newline $:a^1(z)a^2(z)...a^N(z):=:a^1(z)...:a^{N-1}(z)a^N(z):...:$.

Furthermore, as we have seen in the locality axiom of the VOA, two fields $a(z)$ and $b(z)$ are called mutually local if they satisfy $(z-w)^N\{a(z),b(w)\}=0$ for $N\gg 0$.
\subsubsection*{OPE theorem in VOA}\cite{Ka98} (Theorem 2.3)
It has been shown that the operator product expansion (OPE) of two mutually local fields $a(z)$ and $b(w)$ in VOA is given by,
\begin{equation}
    a(z)b(w)=\sum_{j=0}^{N-1}\frac{c_j(w)}{(z-w)^{j+1}}+:a(z)b(w):,
\end{equation}
where $c_j(w) \in End(V)\left[[w,w^{-1}]\right]$. In fact, it has been proved that above OPE product is equivalent to the locality axiom for the $a(z)$ and $b(w)$ fields.
Moreover, the singular part of the OPE is often written as,
\begin{equation}
     a(z)b(w)\sim\sum_{j=0}^{N-1}\frac{c_j(w)}{(z-w)^{j+1}}.\nonumber
\end{equation}
And, the $j$-th product $a(w)*_jb(w)$ of OPE $a(z)b(w)$ is defined as follows,
\begin{eqnarray}
    a(z)b(w)&=&\sum_{j\in \mathbb{Z}}\frac{a(w)*_jb(w)}{(z-w)^{j+1}}=\sum_{j=0}^{N-1}\frac{a(w)*_jb(w)}{(z-w)^{j+1}}+:a(z)b(w):,\nonumber\\
    a(z)b(w)&\sim&\sum_{j=0}^N\frac{a(w)*_jb(w)}{(z-w)^{j+1}}.
\end{eqnarray}
\subsubsection*{The Wick's theorem in VOA}\cite{Ka98} (Theorem 3.3)
Let $a^1(z), ..., a^n(z)$ and $b^1(z), ..., b^m(z)$ be two collections of fields  such that the following properties hold:

1) $\{\{a^i(z)_-,b^j(w)\}, c^k(z)\}=0$ for all $i,j,k,$ and $c=a$ or $b$.

2) $\{a^i(z)_{\pm},b^j(w)_{\pm}\}=0$ for all $i,j$.

let $<a^ib^j>:= \{a^i(z)_-,b^j(w)\}$ denotes the contraction of $a^i(z)$ and $b^j(w)$. Then the following equality holds in the domain $|z|>|w|$:
\begin{equation}
    :a^1(z) ... a^n(z)::b^1(w) ... b^m(w):=\nonumber
\end{equation}
\begin{equation}\label{Wicks theorem VOA}
    \sum_{s=0}^{min(n,m)}\sum_{i_1<...<i_s, j_1\neq...\neq j_s}
    (\pm <a^{i_1}b^{j_1}>...<a^{i_s}b^{j_s}>:a^1(z) ... a^n(z)b^1(w) ... b^m(w):_{(i_1,...,i_s;j_1,..., j_s)}),
\end{equation}
where the sign $\pm$ is obtained by the rule that each permutation of the adjacent odd fields changes the sign and subscript $(i_1,...,i_s;j_1,..., j_s)$ means that the fields $a^{i_1}(z),..., a^{i_s}(z)$ and $b^{j_1}(w),..., b^{j_s}(w)$ are removed.
\subsubsection*{Clifford vertex operator algebra}
The Clifford vertex operator algebra is a vector space $V$ consisting of fermionic states including the vacuum state $|0>=\textbf{1}$, and fermionic vertex operator,
\begin{equation}\label{fermion VO}
Y(\psi_{-\frac{1}{2}}|0>,z)=\sum_{n\in \mathbb{Z}+\frac{1}{2}}\psi_n z^{-n-\frac{1}{2}}:=\psi(z),
\end{equation}
as an odd formal power series with operator modes $\psi_{n}= \frac{1}{2\pi \ii}\oint_{(0)}\zeta^{n-\frac{1}{2}}\psi(z)d\zeta\in End(V)$, satisfying Clifford algebra,
\begin{equation}
\{\psi_n, \psi_m \}= \delta_{n+m,0},\hspace{.2cm} \textit{for}\ n,m\in \mathbb{Z}+\frac{1}{2}.
\end{equation}
The operator modes $\psi_n$ are generators of the Fermionic Fock space, $V$. They act on $V$ as a linear operator such that for any $v\in V$, we have, $\psi_{n}v=0$, for $n\gg 0$ and $\psi_{n}|0>=0$, for $n\ge 0$, and by starting from the vacuum state one can generate basis elements of the fermionic Fock space, e.g. $|\psi>=\psi_{-\frac{1}{2}}|0>=\psi(0)|0>$.
\subsubsection*{Reconstruction theorem for fermion fields} \cite{Ka98} (Theorem 4.5)
As we mentioned, the fermionic Fock space is the vector space $V$ including an even vector, the vacuum $|0>\in V$, and it satisfies the following properties: i) for the fermion fields $\psi(z)$ we have $[T,\psi(z)]=\partial \psi(z)$, ii) $T|0>=0$ and iii) vectors of the following form,
\begin{equation}
    \psi_{-k_n-\frac{1}{2}}\psi_{-k_{n-1}-\frac{1}{2}}...\psi_{-k_2-\frac{1}{2}}\psi_{-k_1-\frac{1}{2}}|0>,\hspace{.2cm}\textit{for}\ k_n>...>k_2>k_1>0,
\end{equation}
span $V$. Then, by reconstruction theorem, for any state in $V$, there is a corresponding field or a vertex operator of the following form,
\begin{equation}\label{reconstruction theorem VOA}
    Y(\psi_{-k_n-\frac{1}{2}}\psi_{-k_{n-1}-\frac{1}{2}}...\psi_{-k_2-\frac{1}{2}}\psi_{-k_1-\frac{1}{2}}|0>,z)=
    :(\partial^{k_n}\psi(z))(\partial^{k_{n-1}}\psi(z))...(\partial^{k_2}\psi(z))(\partial^{k_1}\psi(z)):.
\end{equation}
This defines a unique structure of a vertex algebra on $V$ such that $|0>$ is the vacuum vector, $T$ is the infinitesimal translation operator and $Y(\psi_{-\frac{1}{2}}|0>,z)=\psi(z)$.

\subsubsection*{Conformal vertex algebra for free fermions} \cite{Ka98} (Theorem 4.10)
A conformal vector $\nu \in V$ is an even vector such that the corresponding vertex operator $Y(\nu,z)=L(z)=\sum_{n\in\mathbb{Z}}L_{n}z^{-n-2}$, as an even formal power series, is a Virasoro field of the central charge $c$ with the following properties,
\begin{equation}
L_{-1}=T,\hspace{.2cm}
     \textit{and}\ L_0\  \textit{is diagonalizable on}\ V.
\end{equation}
Thus translation axiom of VOA implies that $[L_{-1},\psi(z)]=\partial_z\psi(z)$.
The vertex algebra $(V, Y, \partial, \textbf{1})$ is called conformal vertex algebra if it has a conformal vector $\nu\in V_2$.
According to the Theorem (4.10) in \cite{Ka98}, $Y(\nu,z)$ is a Virasoro field with central charge $c$ if,
\begin{equation}
    L_{-1}=T,\hspace{.2cm}L_2\nu=\frac{c}{2}|0>, \hspace{.2cm}L_n\nu=0\  \textit{for}\ n>2, \hspace{.2cm}L_0\nu=2\nu.
\end{equation}
Moreover, $\nu\in V_2$ is conformal vector if it satisfies the above properties and the property that $L_0$ is diagonalizable on $V$.

In the fermionic case, $\nu=\frac{1}{2}\psi_{-\frac{3}{2}}\psi_{-\frac{1}{2}}|0>$ is a conformal vector and the Virasoro vertex operator is given by,
\begin{equation}\label{Virasoro VO}
    Y(\frac{1}{2}\psi_{-\frac{3}{2}}\psi_{-\frac{1}{2}}|0>,z)=-\frac{1}{2}:\partial\psi(z)\psi(z): \ =L(z)= \sum_{m\in \mathbb{Z}}L_m z^{-m-2},
\end{equation}
where the modes are given by $L_m=\frac{1}{2\pi \ii}\oint_{(0)}\zeta^{m+1}L(z)d\zeta$. The conformal vector $\nu=\frac{1}{2}\psi_{-\frac{3}{2}}\psi_{-\frac{1}{2}}|0>$ and the Virasoro field $Y(\frac{1}{2}\psi_{-\frac{3}{2}}\psi_{-\frac{1}{2}}|0>,z)$, as an even formal power series, satisfy all the conditions for the conformal vertex algebra with $c=1/2$. 
The explicit form of the Virasoro operator $L_m$ for fermions that satisfies the VOA axioms, in the Sugawara construction for fermionic VOA, can be obtained by using Eqs. (\ref{fermion VO}) and  (\ref{Virasoro VO}), as follows,
\begin{equation}\label{fermionic Virasoro operator}
    L_m=-\frac{1}{2}\sum_{k\in \mathbb{Z}+\frac{1}{2}}(k+\frac{m}{2}):\psi_{m+k}\psi_{-k}: + \frac{1}{16}\delta_{m,0},
\end{equation}
where the normal order is
\begin{equation}
    :\psi_n\psi_m:=\begin{cases}
  \psi_n\psi_m & \hspace{.2cm} n\leq m \\
  -\psi_m\psi_n & \hspace{.2cm} n> m
\end{cases}.
\end{equation}
Straightforward computations shows that the explicit form of the fermionic Virasoro operator (\ref{fermionic Virasoro operator}) satisfies the following commutation relations,
\begin{equation}\label{Virasoro commutation relations}
    [L_m, \psi_n]=-(\frac{1}{2}m+n)\psi_{m+n}, \hspace{.2cm}[L_m,L_n]=(m-n)L_{m+n}+\frac{1}{24}m(m^2-1)\delta_{m+n,0}.
\end{equation}

Fermion fields satisfy the conditions of the Wick's theorem and they are mutually local fields. Thus, the Wick's theorem and the Taylor expansion can be used to obtain the OPE between $L(z)$'s and $\psi(w)$'s,
\begin{eqnarray}
\psi(z)\psi(w)&\sim& \frac{1}{z-w},\nonumber\\
L(z)L(w)&\sim& \frac{1}{4}\frac{I}{(z-w)^4}+\frac{2L(w)}{(z-w)^2}+\frac{\partial L(w)}{(z-w)},\nonumber\\
L(z)\psi(w)&\sim& \frac{1}{2}\frac{\psi(w)}{(z-w)^2}+\frac{\partial_w\psi(w)}{z-w}.
\end{eqnarray}
The commutation relations (\ref{Virasoro commutation relations}) are equivalent to above OPE's.

A fermionic singular state at level two is defined as an state $|\zeta>$ which satisfies,
\begin{equation}
     L_0|\zeta>=(c+2)|\zeta>, \hspace{.2cm}L_n|\zeta>=0,\hspace{.2cm} \textit{for}\ n>0.
\end{equation}
Using the left commutation relation in (\ref{Virasoro commutation relations}), a fermionic singular state at level two can be constructed from the state $\psi_{-\frac{1}{2}}|0>$ as follows,
\begin{equation}\label{singular vector VOA}
    |\zeta>=(L_{-2}+\frac{3}{4}L_{-1}^2)\psi_{-\frac{1}{2}}|0>, \hspace{.2cm}L_0|\zeta>=\frac{5}{2}|\zeta>.
\end{equation}
By using the fermionic representation of $L_m$ in Eq. (\ref{fermionic Virasoro operator}) and the left hand commutation relation in (\ref{Virasoro commutation relations}), one can check that $|\zeta>=0$. In the following, we connect the VOA construction to the fermionic Fock space of fields.
\subsection{Fermionic correlation functions and fermionic Fock space of states}
In previous sections, the construction of the Fock space of fields $\mathcal{F}$, and the Fock space of states, $V$ in the context of VOA are reviewed. In order to connect the Fock space $V$ to the Fock space $\mathcal{F}$, we introduce a map $\chi$ satisfying the following properties.
\subsubsection*{$(\mathcal{F}\leftrightsquigarrow V)$ theorem}There exists a unique collection of maps, $\chi^{(D)}_{(z_1,...z_n)}(v_1\otimes ... \otimes v_n): V^{\otimes n}\rightarrow \mathds{C}$, for $z_1, ..., z_n \in D$, such that following properties are satisfied,

1)
\begin{equation}
    \chi^{(D)}_{(z_1,...,z_n)}(\psi\otimes ... \otimes \psi) = Pf\left(\left[\frac{\sqrt{g'(z_i)}\sqrt{g'(z_j)}}{g(z_i)-g(z_j)}\right]_{i,j=1}^n\right),\nonumber
\end{equation}

2)
\begin{equation}
    \chi^{(D)}_{(z_1,...,z_n)}(v_1\otimes ... \otimes L_{-1}v_i\otimes...\otimes v_n) = \frac{\partial}{\partial z_i}\chi^{(D)}_{(z_1,...,z_n)}(v_1\otimes ... \otimes v_n),\nonumber
\end{equation}

3)
\begin{equation}
    \chi^{(D)}_{(z_1,...,z_n)}(v_1\otimes ... \otimes v_m\otimes v_{m+1}\otimes...\otimes v_n) \sim\nonumber
\end{equation}
\begin{equation}
    \sum_{j=0}^{N-1} \frac{1}{ (z_m-z_{m+1})^{j+1}}\chi^{(D)}_{(z_1,..., \hat{z}_m, z_{m+1},..., z_n)}(v_1\otimes ... \otimes v_m*_jv_{m+1}\otimes...\otimes v_n),
\end{equation}
where $v_m*_jv_{m+1}$ is the $j$-th OPE of the vectors $v_m$ and $v_{m+1}$ and $\hat{z}_m$ is removed.
\subsubsection*{Proof.}
To each vector $v_i\in V$, there is an associated Fock space field $X_i(z_i)\in\mathcal{F}$ and a vertex operator $Y(v_i, z_i)$, then
we define the mapping $\chi^{(D)}_{(z_1,...,z_n)}(v_1\otimes ... \otimes v_m\otimes v_{m+1}\otimes...\otimes v_n)$ for an even $n$ as the correlation function of Fock space fields,
\begin{equation}
    \chi^{(D)}_{(z_1,...,z_n)}(v_1\otimes...\otimes v_n):=<X_1(z_1) ... X_n(z_n)>_D.
\end{equation}
Then, the first axiom follows immediately,
\begin{equation}
    \chi^{(D)}_{(z_1,...,z_n)}(\psi\otimes ... \otimes \psi) = <\psi(z_1) ... \psi(z_n)>_D= Pf\left(\left[\frac{\sqrt{g'(z_i)}\sqrt{g'(z_j)}}{g(z_i)-g(z_j)}\right]_{i,j=1}^n\right).
\end{equation}
By definition, any vector $L_{-1}v_k$, is associated to $L_{-1}X_k(z_k)$ and we have $L_{-1}X_k(z_k)=\partial_{z_k}X_k(z_k)$.
Thus, we obtain the second axiom,
\begin{eqnarray}
    \chi^{(D)}_{(z_1,...,z_n)}(v_1\otimes ... \otimes L_{-1}v_k\otimes...\otimes v_n) &=& <X_1(z_1)...L_{-1}X_k(z_k)... X_n(z_n)>_D \nonumber\\
    &=&<X_1(z_1)... \partial_{z_k}X_k(z_k)... X_n(z_n)>_D\nonumber\\
    &=& \partial_{z_k}<X_1(z_1)... X_n(z_n)>_D\nonumber\\
    &=& \frac{\partial}{\partial z_k}\chi^{(D)}_{(z_1,...,z_n)}(v_1\otimes ... \otimes v_n).
\end{eqnarray}
To prove the third axiom, first let us write the OPE between two Fock space fields,
\begin{equation}
X_m(z_m)X_{m+1}(z_{m+1})\sim\sum_{j=0}^{N-1}\frac{(X_m*_jX_{m+1})(z_{m+1})}{(z_m-z_{m+1})^{j+1}}.
\end{equation}
Then multiply both sides of the above equation by $X_1(z_1)...X_{m-1}(z_{m-1})$ and $X_{m+2}(z_{m+2})...X_n(z_n)$ from left and right, respectively and put them in the correlation function in domain $D$ as follows,
\begin{equation}
<X_1(z_1)...X_{m-1}(z_{m-1})X_m(z_m)X_{m+1}(z_{m+1})X_{m+2}(z_{m+2})...X_n(z_n)>_D\sim\nonumber
\end{equation}
\begin{equation}
\sum_{j=0}^{N-1}\frac{1}{(z_m-z_{m+1})^{j+1}}<X_1(z_1)...X_{m-1}(z_{m-1})(X_m*_jX_{m+1})(z_{m+1})X_{m+2}(z_{m+2})...X_n(z_n)>_D.
\end{equation}
By definition the upper side of the above equation is $\chi^{(D)}_{(z_1,...,z_n)}(v_1\otimes v_2\otimes...\otimes v_n)$. In order to show that,
\begin{equation}
<X_1(z_1)...X_{m-1}(z_{m-1})(X_m*_jX_{m+1})(z_{m+1})X_{m+2}(z_{m+2})...X_n(z_n)>_D=\nonumber
\end{equation}
\begin{equation}
\chi^{(D)}_{(z_1,..., \hat{z}_m, z_{m+1},..., z_n)}(v_1\otimes ... \otimes v_m*_jv_{m+1}\otimes...\otimes v_n),
\end{equation}
we need to show that the corresponding field to the state $v_{m}*_j v_{m+1}$ is $(X_m*_jX_{m+1})(z_{m+1})$. But notice that $v_{m}*_j v_{m+1}$ is the state in the $j$-th product in the OPE of $Y(v_m,z_m)Y(v_{m+1}, z_{m+1})$ in VOA,
\begin{equation}
    Y(v_m,z_{m})Y(v_{m+1},z_{m+1})\sim \sum_{j=0}^{N-1}\frac{(Y(v_m)*_jY(v_{m+1}))(z_{m+1})}{(z_m-z_{m+1})^{j+1}}= \sum_{j=0}^{N-1}\frac{Y(v_{m}*_jv_{m+1},z_{m+1})}{(z_m-z_{m+1})^{j+1}}.
\end{equation}
Thus we need to show that the OPE in VOA and $\mathcal{F}$ are equal. On the half-plane, one can see that the definition of the Fock space fields (\ref{def Fock space field half plane}), and Wick's formula (\ref{Wicks formula half plane}) are the same as definition of fermionic vertex operator in reconstruction theorem (\ref{reconstruction theorem VOA}), and Wick's formula in VOA (\ref{Wicks theorem VOA}), for $a^i(z)=\partial^{k_i}\psi(z)$ and $b^j(z)=\partial^{l_j}\psi(z)$. Thus, the proof of equivalence between the OPE in VOA and OPE in $\mathcal{F}$ is straightforward. However, for domain $D$, from the correspondence between explicit constructions in VOA and Fock space of fields, one could expect the equivalence between the OPE in VOA and OPE in $\mathcal{F}$ but the actual proof remains for future.

This theorem provides a mathematically rigorous approach to the Fock space of the fermionic fields and their correlation functions from the FVOA.
Moreover, we will see in section (5) that the theorem provides us with a rigorous realization of fermionic CFT/$SLE_3$ correspondence at the level of correlation functions and SLE martingale observables.
\section{Fermionic CFT/SLE$_3$ correspondence}
In this section we introduce a rigorous approach to CFT/SLE correspondence, namely the VOA/SLE correspondence. We demonstrate this correspondence in an explicit example, the FVOA/$SLE_3$. We describe the relation between SLE and the scaling limit of statistical lattice model, in a concrete example of the Ising model.
We provide an explicit realization of the CFT/SLE correspondence in the Ising model, first by using the explicit Fock space of fermionic states and its relation to the martingale generators in the case of chordal $SLE_3$. And second, by using the correlation functions of fermions on the upper-half plane and their differential equations, Ward identity and null field differential equation, we show that the correlation functions of fermionic Fock space fields on domain $D$ produce chordal $n-SLE_3$ martingale observables.

Stochastic Loewner evolution is a conformally invariant stochastic process satisfying domain Markov property and domain conformal invariance, introduced in \cite{Sch00}. For further details see \cite{La08}.
Let us briefly describe the simplest case of chordal SLE curves in domain $(D, a, b)$. The chordal $SLE_\kappa$, with $\kappa\geq 0$ is conformally invariant random curve processes in domain $D$ from $a$ to $b$, two boundary points. It is described by the Loewner equation with a one-dimensional Brownian motion $B_t$ as a driving force.
Let $g_t(z)$ for $z\in D$ and $t<\tau_z\in (0, \infty]$ be the solution of the following equation,
\begin{equation}
    \frac{dg_t(z)}{dt} = \frac{2}{g_t(z)-\sqrt{\kappa}B_t},\hspace{.5cm} B_0=0,\nonumber
\end{equation}
where $g_0: (D, a, b)\to (\mathbb{H}, 0, \infty)$ is a conformal map.  Then, the function $h_t(z):=g_t(z)-\sqrt{\kappa}B_t$ for all $t$ is a conformal map, $h_t(z):D_t\to \mathbb{H}$ from domain $D_t:=\{z\in D:t<\tau_z\}$ onto $\mathbb{H}$, where $h_t(z)$ at $t= \tau_z$ maps the tip of the curve to zero; $\lim_{t\to \tau_z}h_t(z)=0$. The chordal $SLE_\kappa$ curve $\gamma$ is defined by $\gamma(t):=\lim_{z\to 0}g_t^{-1}(z+\sqrt{\kappa}B_t)$.

\subsubsection*{Fermionic realization of $CFT/SLE$ in Ising model}
The relation between CFT and SLE have been studied from different perspectives, for reviews on the subject see \cite{BaBe06}, \cite{Car05}, \cite{Gr06} and \cite{Kon03}. In one perspective, the relation between the interfaces in the scaling limit of critical lattice models and the field theory describing that limit is a natural question to study. The interfaces are classified by chordal SLE curves which are characterized by a parameter $\kappa$. Moreover, the scaling limits of critical lattice models are usually described by conformal field theories which are characterized by a central charge $c$. This classifying number, determines the universality class of the scaling limit of the different lattice models such as Ising model. 

In general, a CFT corresponds to a chordal SLE if the parameters of CFT, $c$ and $h$, and the parameter of chordal SLE, $\kappa$ satisfy,
\begin{equation}
    c_\kappa= \frac{(3\kappa-8)(6-\kappa)}{2\kappa}, \hspace{.5cm} h=\frac{6-\kappa}{2\kappa}.
\end{equation}
In our example the values of parameters, $c=\frac{1}{2}, h=\frac{1}{2}$ and $\kappa=3$, satisfy the above equations.

Although, there is a common belief that a CFT with $c=\frac{1}{2}$ such as $m=3$ minimal model describes the scaling limit of the critical Ising model \cite{DMS96} (chapter 12), but there were no systematic approach proposed to CFT based on probability theory until recently. We study towards an algebraic construction of a conformal quantum field theory of free fermions for Ising model based on probability theory, SLE.
In fact, in the Ising model example, we apply $SLE_3$ and FVOA to construct and study fermionic CFT of the scaling limit of the Ising model, rigorously.

On the one hand, we studied the rigorous scaling limit of Ising model and its fermionic correlation functions which lead to a fermionic conformal field theory. As we have seen in previous chapter, this is composed of algebraic Fock space of states, local Fock space fields, their correlation functions and differential equations. On the other hand, $SLE_3$ curves and observables appear in the scaling limit of the Ising model at criticality. In this section, we study the relation between these two distinct scaling limit of Ising model in different aspects.

First, we study the algebraic operator formalism of fermionic Fock space of fields and states in CFT and VOA and its relation to martingale generators of $SLE_3$. Second, we explain how a certain fermionic observable of $2d$ critical Ising model, which is obtained from the scaling limit of Ising fermion correlation functions, is related to a martingale observable of $SLE_3$.

The first explicit realization of fermionic $CFT/SLE_3$ in the case of Ising model is the explicit form of the boundary conditions changing (B.C.C.) operator, the Ising fermion field $\psi(z)$, that is obtained from the scaling limit of the Ising lattice fermion operator. This is an operator that changes the boundary conditions on the lattice row configurations from minus (plus) to plus (minus) at the insertion point. As we have shown in previous section, the state $\psi_{-\frac{1}{2}}|0>$ is a primary state and degenerate at level two. Therefore, this state can be considered as an explicit B.C.C. state for the chordal $SLE_3$ curve. In fact, the field $\psi(z)$ is a boundary operator, inserted in a boundary point $z$, the starting point of the $SLE_3$ curve.
\subsection{$SLE_3$ and interfaces in Ising model}
SLE has been used widely in the study of lattice models such as Ising model, percolation, etc. The scaling limit of lattice model interfaces at criticality is described by the SLE curves.
Let us start with an intuitive picture of how the SLE curve emerges in the Ising model. A chordal $SLE_3$ curve appears as the scaling limit of discrete interface or boundary between two boundary points where the boundary conditions change. The interface separates two clusters of plus and minus spins in the Ising model. The claim is that in the scaling limit, the interface converges to the chordal $SLE_3$ curve. The convergence of the interfaces in the Ising model to the $SLE_3$ curves has been proved recently in \cite{CDHKS12}.

The probability measure of the spin configurations in the Ising model induces a probability measure on the interfaces, in the following sense. Consider a domain $D$ in complex plane with two boundary points $a,b$. Approximate the domain and boundary points by square lattice domain $D_n= \frac{1}{n}\mathds{Z}^2\cap D$ and $a_n, b_n$, respectively, and define a probability measure on interfaces in this domain, $P_n(D_n, a_n, b_n)$.
In the limit $n\rightarrow \infty$, $P_n$ converges to $\mu(D, a, b)$ which is the law of chordal $SLE_3$ in the domain $D$ from $a$ to $b$. This procedure can be easily generalized for the the case of several interfaces.

Let us consider the general case of chordal $n$-$SLE_3$, where there are $2n$ starting points and arbitrary number of bulk points. The general picture of the chordal $n$-$SLE_3$ consists of several interface curves, growing by Loewner chain, a collection of $g_t$, with random driving force, which connect the boundary points in the critical Ising model. Moreover, in order to define a chordal $n$-$SLE_3$ we have to write down the Loewner equation in $\mathbb{H}$ with explicit conditions. By using a hydrodynamically normalized conformal map $g_t$, we can uniformize the complement of the several $SLE_3$ curves, labeled by integer $i$ with the starting point $X_i$. This conformal map satisfies the Loewner equation,
\begin{eqnarray}
  d g_t(z)&=&\sum_{i=1}^{2n} \frac{2v_t^i dt}{g_t(z)-X_t^i},\nonumber \\
  g_0(z) &=& z,
\end{eqnarray}
where $v_t^i$ is the growth speed of $i$-th curve which can be set to $ v_t^i=1$ and $X_t^i$ are the images of the tips of the curves under the map $g_t$.

In order to define a martingale observable from the fermionic correlation functions for chordal $n$-$SLE_3$ we claim that $X^i_t$ should satisfy,
\begin{equation}
    dX_t^i=\sqrt{3}dB_t^i+3(\partial_{x_i}\log Z_{n-SLE_3}^{\mathbb{H}})dt+\sum_{l\neq i}\frac{2 dt}{X_t^i-X_t^l},
\end{equation}
where $dB_t^i$ are $2n$ independent Brownian motions, $Z_{n-SLE_3}^{\mathbb{H}}$ is called the $SLE_3$ partition function which will be defined in the following section and initial conditions are $X_0^i=X_i$, and they are ordered as $X_1<X_2<...<X_{2n}$. Our claim will be proved in section (5.3). 
\subsubsection*{$SLE_3$ partition function}
In the simplest case, a partition function of the chordal $SLE_3$ on the domain $D$ with two boundary points $a,b \in \partial D$ is defined in terms of the two-point fermionic correlation function,
\begin{equation}
Z_{SLE_3}^{D}=\chi_{a,b}^D(\psi\otimes\psi)=<\psi(a)\psi(b)>_{D}\sim \frac{1}{a-b}.
\end{equation}

In the case of chordal $n$-$SLE_3$ the partition function is given by the correlation functions of fermions on domain $D$ and it has the Pfafffian structure,
\begin{equation}
    Z_{n-SLE_3}^D=\chi_{x_1,...,x_{2n}}^D(\psi\otimes...\otimes\psi)= <\prod_{i=1}^{2n} \psi(x_i)>_D=Pf\left(\left[\frac{\sqrt{\varphi'(x_i)}\sqrt{\varphi'(x_j)}}{\varphi(x_i)-\varphi(x_j)}\right]_{i,j=1}^{2n}\right),
\end{equation}
where operators $\psi$'s are inserted at boundary points, $x_1, ..., x_{2n}\in D$ at B.C.C. points and $\varphi: D\to \mathbb{H}$ is a conformal map.
It can be checked that the chordal $n$-$SLE_3$ partition function on the half-plane, $Z_{n-SLE_3}^{\mathbb{H}}$ as a CFT correlation function of fermion fields with Pfaffian form satisfies the homogeneity and scaling equations,
\begin{equation}
    \sum_i\frac{\partial}{\partial x_i}Z_{n-SLE_3}^{\mathbb{H}}=0, \hspace{.3cm}\sum_i(x_i\frac{\partial}{\partial x_i}+\frac{1}{2})Z_{n-SLE_3}^{\mathbb{H}}=0, \hspace{.3cm}
    \sum_i(x_i^2\frac{\partial}{\partial x_i}+x_i)Z_{n-SLE_3}^{\mathbb{H}}=0,
\end{equation}
and the null field differential equation,
\begin{equation}\label{null field differential equation for SLE}
    \left[\frac{3}{4}\frac{\partial^2}{\partial x_i^2}+
    \sum_{l\neq i}\left[\frac{1}{x_l-x_i}\frac{\partial}{\partial x_l}-\frac{1/2}{(x_l-x_i)^2}\right]\right]Z_{n-SLE_3}^{\mathbb{H}}=0,
\end{equation}
for $i=1, ..., 2n$.
As it will be explained in the following section, this partition function can be used to define the local martingales in chordal $n$-$SLE_3$.
\subsection{$SLE_3$ Martingale generators and fermionic Fock space of states}
In this section we study the fermionic Fock space of states of the chordal $SLE_3$ curves. By using the Clifford VOA, we want to construct explicitly the Fock space of states of chordal $SLE_3$ curves, or the chordal $SLE_3$ martingale generators, in domain $H_t :=\mathbb{H}\setminus \gamma[0,t]$.

Let us review construction of the operator formalism in SLE. It is known that, to any formal power series in particular function $f\in z+\mathbb{C}[[z^{-1}]]$ of the form $f(z)=z+\sum_{m\leq -1}f_m z^{1+m}$, one can associate an operator $G_f$,
\begin{equation}
  G_f \in \overline{\mathcal{U}(\mathfrak{vir}_-)}=\prod_{d=0}^{\infty}\mathcal{U}(\mathfrak{vir}_-)_d,
\end{equation}
where the algebra $\overline{\mathcal{U}(\mathfrak{vir}_-)}$ is a completion of the universal enveloping algebra $\mathcal{U}(\mathfrak{vir}_-)=\bigoplus_{d=0}^\infty\mathcal{U}(\mathfrak{vir}_-)_d$ of the Virasoro subalgebra $\mathfrak{vir}_-$ generated by $L_n (n<0)$, \cite{BaBe04}.

As we observed, the explicit fermionic Fock space of states $V$ is constructed from the Clifford VOA in section (4.1) and $V$ consists of basis vectors of the form
$\psi_{-k_n-\frac{1}{2}}\psi_{-k_{n-1}-\frac{1}{2}}...\psi_{-k_2-\frac{1}{2}}\psi_{-k_1-\frac{1}{2}}|0>$. The space of fermionic descendant states $\mathcal{M}$, and space of descendant Fock space fields $\mathcal{N}$, can be constructed by the action of Virasoro generators on the state $|\psi>= \psi_{-\frac{1}{2}}|0>$ and its corresponding field $\psi(z)$, as follows,
\begin{equation}
    \mathcal{M}=\mathcal{U}(\mathfrak{vir}_-)|\psi> \subset V,\hspace{.5cm}  \mathcal{N}=\mathcal{U}(\mathfrak{vir}_-)\psi (z) \subset \mathcal{F}.
\end{equation}
Let us define the intertwining operator $G_{h_t}$. This is an operator, associated to the conformal map $h_t= g_t- \sqrt{3}B_t$ where $g_t:H_t\to \mathbb{H}$, between domains $\mathbb{H}$ and $H_t$.
By using the intertwining operator $G_{h_t}$, we can construct the $SLE_3$ martingale generators and their Fock space in the domain $H_t$.
The Fock space of $SLE_3$ curves and the Fock space of fields in $H_t$ are obtained from the corresponding Fock spaces in $\mathbb{H}$ by means of the action of the transformation operator, $G_{h_t}|\psi>$, and $G_{h_t}^{-1}\psi(z)G_{h_t}$, as follows,
\begin{equation}
    \mathcal{M}_t= G_{h_t}\mathcal{U}(\mathfrak{vir})|\psi>,\hspace{.5cm} \mathcal{N}_t= G_{h_t}^{-1}(\mathcal{U}(\mathfrak{vir})\psi(z))G_{h_t},
\end{equation}
where the elements of $\mathcal{M}_t$ are of the form $G_{h_t}\psi_{-k_n-\frac{1}{2}}\psi_{-k_{n-1}-\frac{1}{2}}...\psi_{-k_2-\frac{1}{2}}\psi_{-k_1-\frac{1}{2}}|0>$, and the elements of $\mathcal{N}_t$ have the following form, $G_{h_t}^{-1}:(\partial^{k_n}\psi(z))(\partial^{k_{n-1}}\psi(z))...(\partial^{k_2}\psi(z))(\partial^{k_1}\psi(z)):G_{h_t}$. 
\vspace{.2cm}
Using the grading of $V=\bigoplus_{h\in \frac{1}{2}\mathbb{N}} V_h$, for $\forall h$ we can explicitly write the vector valued graded martingale generators $M_h$ of the chordal $SLE_3$ as follows,
\begin{equation}
   \frac{1}{Z}G_{h_t}|\psi> = \sum_{h\in \frac{1}{2}\mathds{N}}M_h\in \bar V,
\end{equation}
where $Z$ is the chordal $SLE_3$ partition function on the half plane and in fact $Z=<\psi(\infty)\psi(0)>_{\mathbb{H}}=1$, and $\bar V= \prod_{h\in \frac{1}{2}\mathds{N}}V_h$ is the completion of $V$.

As we observed in Eq. (\ref{singular vector VOA}), the state $|\psi>\in V$ is a primary state degenerate at level two with singular descendant state at level two, $(L_{-2}+\frac{3}{4}L_{-1}^2)|\psi>=0$. Using that, it can be shown that the state $G_{h_t}|\psi>$ is a local martingale of chordal $SLE_3$, which means that $<v|G_{h_t}|\psi>$ is conserved in mean for any $<v|$,
\begin{equation}
    \mathbb{E}[<v|G_{h_t}|\psi>|\{G_{h_u}\}_{u\leq s}]=<v|G_{h_s}|\psi>,
\end{equation}
for $t\geq s$.

In order to prove the local martingale property for $G_{h_t}|\psi>$, we need to show that the drift term in It\^{o} derivative of this state vanishes. Since $G_{h_t}$ is an intertwining operator corresponding to the conformal map $h_t$, it satisfies the It\^{o} differential equation, see section (5.3.3) in \cite{BaBe06},
\begin{equation}
    G_{h_t}^{-1}dG_{h_t}= dt(-2L_{-2}+\frac{3}{2}L_{-1}^2)-d\xi_t L_{-1}.
\end{equation}
Then, if we apply both sides of above equation to the state $|\psi>$ and use the level two singular vector equation then we obtain,
\begin{equation}
    dG_{h_t}|\psi>=G_{h_t}L_{-1}|\psi>d\xi_t.
\end{equation}
The above equation makes perfect sense in $\bar V$ and shows that the drift term in the It\^{o} derivative vanishes and thus $G_{h_t}|\psi>$ is a local martingale of chordal $SLE_3$.

Since that is a local martingale, all the fermionic correlation functions of CFT in the domain $H_t$ which are $SLE_3$ observable on the domain $H_t$, and are constructed from the vector $G_{h_t}|\psi>$ will be local martingale observables of chordal $SLE_3$. Therefore, $G_{h_t}|\psi>$ is called a generating function of local martingales of chordal $SLE_3$.

In general, the scaling limit of the interfaces in Ising model are related to the scaling limit of correlation functions of local operators and fields in the Ising model. We will elaborate on this point in the following section.
\subsection{$SLE_3$ Martingale observables and fermion correlation functions}
Let us first describe the martingale observables. It is easy to show that an observable is a local martingale if the drift term in its It\^{o} derivative vanishes. The general claim of this part is that a large collection of $SLE_3$ martingale observables can be constructed systematically by using the rigorously constructed correlation functions of fermionic Fock space fields in $(\mathcal{F}\leftrightsquigarrow V)$ theorem.

In this section, we present an explicit form of some chordal $n-SLE_3$ martingale observables as the correlation functions of free fermion fields and descendant fermionic Fock space fields. Furthermore, we will see that basic $SLE_3$ martingale observables, i.e. observables obtained from the correlation functions of free fermion fields, have an explicit Pfaffian structure.
We study the relations between fermionic null field differential equation (\ref{null field differential equation}) and vanishing of the drift term in the It\^{o} formula for the $SLE_3$ martingale observables.

Let us construct an example of chordal $n$-$SLE_3$ martingale observable, see \cite{BBK05}. Consider an operator of the form $\mathcal{O}=\prod_{k=1}^m\psi(W_k)$. We claim that a martingale observable of this operator in the domain $H_t$, where the $2n$ $SLE_3$ curves are removed from $\mathbb{H}$ and the tips of the $SLE_3$ curves are $\gamma_1, ..., \gamma_{2n}$, is
\begin{equation}\label{local martingale observable}
<\mathcal{O}>_{H_t}=\frac{1}{Z^{H_t}_{n-SLE_3}}\chi^{H_t}_{\gamma_1, ..., \gamma_{2n}; W_1, ..., W_{m}}((\psi\otimes ... \otimes\psi)\otimes(\psi\otimes ... \otimes\psi)),
\end{equation}
where $Z^{H_t}_{n-SLE_3}=\chi^{H_t}_{\gamma_1, ..., \gamma_{2n}}(\psi\otimes ... \otimes\psi)$ is a partition function of chordal $n$-$SLE_3$ in domain $H_t$.
In this case, the explicit form of the correlation function $\chi^{H_t}_{\gamma_1, ..., \gamma_{2n}; W_1, ..., W_{m}}((\psi\otimes ... \otimes\psi)\otimes(\psi\otimes ... \otimes\psi))$ and the partition function are known as Pfaffian formulas. In the following, we prove the claim that Eq. (\ref{local martingale observable}) gives a local martingale observable of the chordal $n$-$SLE_3$.

The observable $<\mathcal{O}>_{H_t}$ can be transformed to $\mathbb{H}$ by using the mapping $g_t: H_t\to \mathbb{H}$, as follows,
\begin{equation}\label{SLE observable on the mapped domain}
    <\mathcal{O}>_{H_t}= \frac{1}{Z^{\mathbb{H}}_{n-SLE_3}}J_t\chi^{\mathbb{H}}_{x_1, ..., x_{2n}; w_1, ..., w_{m}}((\psi\otimes ... \otimes\psi)\otimes(\psi\otimes ... \otimes\psi)),
\end{equation}
where $Z^{\mathbb{H}}_{n-SLE_3}=\chi^{\mathbb{H}}_{x_1, ..., x_{2n}}(\psi\otimes ... \otimes\psi)$, $x_i=g_t(\gamma_i)$, $w_k= g_t(W_k)$ and the Jacobian is $J_t=\prod_{k=1}^m g_t'^{\frac{1}{2}}(W_k).$
Above chordal $n$-$SLE_3$ observable is a local martingale if the drift term in the It\^{o} derivative of the observable vanishes. To show that the drift term vanishes, we need It\^{o} formula for the $\psi(X^i_t)$, which is $d\psi(X^i_t)=\psi'(X^i_t)dX^i_t+\frac{3}{2}\psi''(X^i_t)dt$, and Loewner equation for $g_t(W_k)$ and its derivative with respect to $W_k$ which lead to,
\begin{equation}
d(\psi(w_k)w_k'^{\frac{1}{2}})= w_k'^{\frac{1}{2}}\sum_i 2dt\left(\frac{\psi'(w_k)}{w_k-X^i_t}-\frac{\frac{1}{2}\psi(w_k)}{(w_k-X^i_t)^2}\right).
\end{equation}
In addition, we need the null field differential equation for the chordal $n$-$SLE_3$ partition function on $\mathbb{H}$, Eq. (\ref{null field differential equation for SLE}), and the following equation for the correlation function of fermions on $\mathbb{H}$,
\begin{equation}\label{null diff equation for VOA}
    D\chi^{\mathbb{H}}_{x_1, ..., x_{2n}; w_1, ..., w_{m}}((\psi\otimes ... \otimes\psi)\otimes(\psi\otimes ... \otimes\psi))=0,
\end{equation}
where operator $D$ is
\begin{eqnarray}
    D=(\frac{3}{4}\frac{\partial^2}{\partial x_i^2}
   &+&\sum_{l\neq i}\left[\frac{1}{x_l-x_i}\frac{\partial}{\partial x_l}-\frac{1/2}{(x_l-x_i)^2}\right]\nonumber\\
    &+&\sum_{k=1}^m\left[\frac{1}{w_k-x_i}\frac{\partial}{\partial w_k}-\frac{1/2}{(w_k-x_i)^2}\right]).
\end{eqnarray}
Then, let us write the It\^{o} derivative of the numerator of Eq. (\ref{SLE observable on the mapped domain}),
\begin{equation}
    d(J_t\chi^{\mathbb{H}}_{x_1, ..., x_{2n}; w_1, ..., w_{m}}((\psi\otimes ... \otimes\psi)\otimes(\psi\otimes ... \otimes\psi)))=\nonumber
\end{equation}
\begin{equation}
    J_t\left[\sum_idX^i_t\partial_{x_i}+\sum_idt\left(\frac{3}{4}\frac{\partial^2}{\partial x_i^2}
   +\sum_{k=1}^m\left[\frac{1}{w_k-x_i}\frac{\partial}{\partial w_k}-\frac{1/2}{(w_k-x_i)^2}\right]\right)\right]\nonumber
\end{equation}
\begin{equation}
    \chi^{\mathbb{H}}_{x_1, ..., x_{2n}; w_1, ..., w_{m}}((\psi\otimes ... \otimes\psi)\otimes(\psi\otimes ... \otimes\psi)).
\end{equation}
Then, by using the null differential Eq. (\ref{null diff equation for VOA}), the It\^{o} derivative becomes,
\begin{equation}
    d(J_t\chi^{\mathbb{H}}_{x_1, ..., x_{2n}; w_1, ..., w_{m}}((\psi\otimes ... \otimes\psi)\otimes(\psi\otimes ... \otimes\psi)))=\nonumber
\end{equation}
\begin{equation}
    J_t\left[\sum_idX^i_t\partial_{x_i}-\sum_i2dt\left( \sum_{l\neq i}\left[\frac{1}{x_l-x_i}\frac{\partial}{\partial x_l}-\frac{1/2}{(x_l-x_i)^2}\right]\right)\right]\nonumber
\end{equation}
\begin{equation}
    \chi^{\mathbb{H}}_{x_1, ..., x_{2n}; w_1, ..., w_{m}}((\psi\otimes ... \otimes\psi)\otimes(\psi\otimes ... \otimes\psi)).
\end{equation}
As an special case, the above equation implies that,
\begin{equation}
    dZ_{n-SLE_3}^{\mathbb{H}}=\left[\sum_idX^i_t\partial_{x_i}-\sum_i2dt\left( \sum_{l\neq i}\left[\frac{1}{x_l-x_i}\frac{\partial}{\partial x_l}-\frac{1/2}{(x_l-x_i)^2}\right]\right)\right]Z_{n-SLE_3}^{\mathbb{H}}.
\end{equation}
Finally, by using the above equations the It\^{o} derivative of $<\mathcal{O}>_{H_t}$ becomes,
\begin{equation}
    d<\mathcal{O}>_{H_t}=J_t\sum_i\left[dX^i_t-(3\partial_{x_i}\log Z^{\mathbb{H}}_{n-SLE_3}+2\sum_{l\neq i}\frac{1}{x_i-x_l}) dt\right]\partial_{x_i}\nonumber
\end{equation}
\begin{equation}
    (\frac{1}{Z^{\mathbb{H}}_{n-SLE_3}}\chi^{\mathbb{H}}_{x_1, ..., x_{2n}; w_1, ..., w_{m}}((\psi\otimes ... \otimes\psi)\otimes(\psi\otimes ... \otimes\psi))).
\end{equation}
Thus, we observe that if the process $X^i_t$ satisfies,
\begin{equation}\label{condition for martingale observable}
    dX^i_t=\sqrt{3}dB^i_t+(3\partial_{x_i}\log Z^{\mathbb{H}}_{n-SLE_3}+2\sum_{l\neq i}\frac{1}{x_i-x_l}) dt,
\end{equation}
then the drift term in It\^{o} formula for $<\mathcal{O}>_{H_t}$ vanishes and we have,
\begin{equation}
    d<\mathcal{O}>_{H_t}=\sqrt{3}J_t\sum_idB^i_t\partial_{x_i}[\frac{1}{Z^{\mathbb{H}}_{n-SLE_3}}\chi^{\mathbb{H}}_{x_1, ..., x_{2n}; w_1, ..., w_{m}}((\psi\otimes ... \otimes\psi)\otimes(\psi\otimes ... \otimes\psi))].
\end{equation}

To summarize, we observe that $\frac{1}{Z^{H_t}_{n-SLE_3}}\chi^{H_t}_{\gamma_1, ..., \gamma_{2n}; W_1, ..., W_{m}}((\psi\otimes ... \otimes\psi)\otimes(\psi\otimes ... \otimes\psi))$, with the condition (\ref{condition for martingale observable}), is a martingale observable of chordal $n$-$SLE_3$.

Without explicit calculations, but because of the fact that $<v|G_{h_t}|\psi>$ (for any arbitrary $<v|$ which can be obtained from the action of descendant fields on the $<\psi|$) are local martingales, we claim that the correlation functions of arbitrary Fock space fields can be used to write the following expression as a chordal $n$-$SLE_3$ martingale observable,
\begin{equation}\label{general martingale observable}
    \frac{1}{Z^{H_t}_{n-SLE_3}}\chi^{H_t}_{\gamma_1, ..., \gamma_{2n}; W_1, ..., W_{m}}((\psi\otimes ... \otimes\psi)\otimes(v_1\otimes ... \otimes v_m)),
\end{equation}
where $v_i$'s are arbitrary vectors in Fock space. Moreover, we claim that all the $SLE_3$ observables obtained from the fermionic correlation functions are reducible to the basic $SLE_3$ observables by using the generalization of Eq. (\ref{generalized Ward identity}).
\section{Conclusions}
In this article we have studied a concrete and explicit realization of the $CFT/SLE$ correspondence in the case of Ising model. We obtained the correlation functions of free fermionic fields on domain $D$ by taking the scaling limit of the lattice correlation functions of the Ising free fermions. These results are obtained by using the rigorous methods of discrete holomorphicity and Riemann boundary value problem introduced in \cite{Hon10a}. We investigated the algebraic and analytic fermionic conformal field theory by studying the correlation functions of fermion fields and differential equations that they satisfy such as Ward identity and singular vector differential equations. Moreover, we developed the algebraic aspects of the fermionic Fock space of states and fermionic vertex operator algebra and especially we found a mapping between the Fock space of states and the correlation functions of the fermionic fields which respects the conformal structure of the theory. The relation between these results and the probability theory of martingale generators and observables in chordal $SLE_3$ are studied and we have worked out all these relations explicitly in a concrete example of the Ising model.

A possible direction for further studies is to complete the proof of the ($\mathcal{F}\leftrightsquigarrow V$) theorem. Furthermore, explicit proof of the martingale observable (\ref{general martingale observable}) remains for future studies.

\thanks{\textbf{Acknowledgements:} I would like to thank Kalle Kyt\"{o}l\"{a} for useful and interesting discussions and comments. This work was supported by the Academy of Finland.}

{}
\vspace{10pt}
\address{Department of Mathematics and Statistics, P.O. Box 68, FIN\textendash{}00014
University of Helsinki, Finland}\\
\email{\textit{E-mail address}: seyedali.zahabi@helsinki.fi}
\end{document}